\documentclass[journal,12pt,onecolumn,draftclsnofoot,]{IEEEtran}
\usepackage{algorithm} 
\usepackage{algpseudocode}
\usepackage{amsmath,amsfonts,amssymb}
\usepackage{amsthm}
\usepackage{bm}
\usepackage{booktabs}
\usepackage{caption}
\usepackage[scale=2]{ccicons}
\usepackage{cite}
\usepackage{color}
\usepackage{etoolbox}
\usepackage{graphicx}
\usepackage{lettrine}
\usepackage{mathtools}
\usepackage{multirow}
\usepackage{pgfplots}
\usepackage{ragged2e}
\usepackage{rotating}
\usepackage{setspace}
\usepackage{subfigure}
\usepackage{url}
\usepackage{xspace}
\usepackage{tikz}
\usetikzlibrary{shapes,arrows}
\PassOptionsToPackage{bookmarks=false}{hyperref}

\newtheorem{mydef}{Definition}
\newtheorem{prop}{Proposition}

\begin{document}
	\title{Ruin Theory for Energy-Efficient Resource Allocation in UAV-assisted Cellular Networks}
	
	%
	
	\author{
		\IEEEauthorblockN{Aunas~Manzoor,
			Kitae~Kim,
			Shashi~Raj~Pandey,
			S.~M.~Ahsan~Kazmi,
			Nguyen~H.~Tran,~\IEEEmembership{Senior~Member,~IEEE},
			Walid~Saad,~\IEEEmembership{Fellow,~IEEE},				
			and~Choong~Seon~Hong,~\IEEEmembership{Senior~Member,~IEEE}}
		
		\thanks{A. Manzoor, Kitae Kim, S. R. Pandey, and C. S. Hong  are with the Department of Computer Science and Engineering, Kyung Hee University, Yongin 446-701, South Korea (e-mail: \{aunasmanzoor; glideslope; shashiraj; cshong\}@khu.ac.kr).}
		\thanks{S. M. Ahsan Kazmi is with Networks and Blockchain Lab, Institute of Secure and Cyber Physical System, Innopolis University, Tatarstan 420500, Russia, and also with the Department of Computer Science and Engineering, Kyung Hee University, Yongin 446-701, South Korea (e-mail: a.kazmi@innopolis.ru).}
		\thanks{Nguyen H. Tran is with the School of Computer Science, The University of Sydney, Sydney, NSW 2006, Australia (e-mail: nguyen.tran@sydney.edu.au).}
		\thanks{W. Saad is with Wireless@VT, Bradley Department of Electrical and Computer Engineering, Virginia Tech, Blacksburg, VA 24061 USA, and also with the Department of Computer Science and Engineering, Kyung Hee University, Yongin 446-701, South Korea (e-mail: walids@vt.edu).}
	}
	
	
	%
	\pagenumbering{gobble}	
	\maketitle
	\begin{abstract}
		Unmanned aerial vehicles (UAVs) can provide an effective solution for improving the coverage, capacity, and the overall performance of terrestrial wireless cellular networks. In particular, UAV-assisted cellular networks can meet the stringent performance requirements of the fifth generation new radio (5G NR) applications. In this paper, the problem of energy-efficient resource allocation in UAV-assisted cellular networks is studied under the reliability and latency constraints of 5G NR applications. The framework of ruin theory is employed to allow solar-powered UAVs to capture the dynamics of harvested and consumed energies. First, the surplus power of every UAV is modeled, and then it is used to compute the probability of ruin of the UAVs. The probability of ruin denotes the vulnerability of draining out the power of a UAV. Next, the probability of ruin is used for efficient user association with each UAV. Then, power allocation for 5G NR applications is performed to maximize the achievable network rate using the water-filling approach. Simulation results demonstrate that the proposed ruin-based scheme can enhance the flight duration up to 61\% and the number of served users in a UAV flight by up to 58\%, compared to a baseline SINR-based scheme.                  
	\end{abstract}
	
	\vspace{0.5 in}
	
	\IEEEpeerreviewmaketitle
	\section{Introduction}
	
	
	\IEEEPARstart{T}{he}  use of \emph{unmanned aerial vehicles (UAVs)} can enable a wide range of smart city applications, ranging from drone delivery to surveillance and monitoring \cite{1677946,mozaffari2019beyond,mozaffari2018tutorial}. Recently, the use of UAVs has greatly increased in wireless-networking applications to provide coverage and capacity enhancement to the ground wireless networks \cite{saad2019vision}. The flexibility, autonomy, and ease of deployment of UAVs render them suitable to be part of the future wireless networks. In wireless networking applications, UAVs can have various roles that range from flying base stations (BSs) (\hspace{1sp}\cite{mozaffari2018tutorial}) to backhaul nodes (\hspace{1sp}\cite{kalantari2017backhaul}) and users of the cellular network (\hspace{1sp}\cite{challita2018cellular,mozaffari2018tutorial,saadmobsky}). Therefore, by leveraging line-of-sight (LoS) communication at high altitudes as well as the dynamic placement of UAVs at desired locations and within a required time \cite{chen2017optimal}, the use of UAVs as flying BSs can play a significant role in boosting the capacity of cellular networks \cite{sharma2016uav}. For example, in \cite{uavhetnet}, the authors proposed an UAV-assisted heterogeneous cellular network (HetNet) to meet the communication demands in emergencies for public safety. The work in \cite{saadtransport} used optimal transport theory to enable UAVs to provide communication services to ground users while optimizing their flight time. Moreover, the authors of \cite{khoshkholgh2019coverage} proposed an efficient UAV BSs in coexistence with a terrestrial network. In particular, when wireless connectivity is needed in difficult and costly deployment locations, UAVs can provide low-cost and low-power alternatives and complement conventional small cell BSs (SBSs). For instance, a joint positioning and user association problem was addressed in \cite{uavhetnet1}, where UAVs were used as a replacement of terrestrial SBSs. Moreover, UAVs are a very promising solution to the problem of connectivity in occasionally crowded areas, such as stadiums or open-air shows. Owing to these benefits of UAVs in wireless communications, it is envisioned that future wireless cellular networks \cite{saad2019vision} will be \emph{UAV-assisted} because they can complement their terrestrial infrastructure with flying UAV BSs.  Meanwhile, to design efficient UAV-assisted cellular networks, it is necessary to address various challenges that range from network modeling to optimization and resource management \cite{mozaffari2018tutorial}.  
	
	The use of UAVs is particularly meaningful for delivering 5G new radio (NR) applications \cite{bennis2018ultrareliable}. Since UAVs have already been used for data collection from Internet of Things (IoT) networks \cite{saadiot}, they can perform well for massive machine type communication (mMTC). Moreover, to meet the reliability and latency demands of ultra-reliable and low-latency communication (URLLC) applications, the LoS communication, and optimal positioning features of UAVs can be promising. Recently, the use of UAVs has been proposed in delay-sensitive and mission critical applications \cite{fadlullah2016dynamic,ren2019achievable,pan2019joint}. The authors in \cite{fadlullah2016dynamic} proposed a dynamic trajectory control algorithm to optimize the delay and throughput in the UAV-aided networks. In \cite{ren2019achievable}, the authors developed a URLLC channel model to ensure the delay-sensitive delivery of critical control information from the ground station to the UAV. The work in \cite{pan2019joint} used a URLLC-enabled UAV to develop a relay system for the delay-sensitive and ultra-reliable communication. However, most of these prior works focused on the use of UAVs to incorporate mMTC or URLLC services, rather than consolidating all of the 5G services. Therefore, we propose the UAV-assisted cellular networks to serve the cellular users while incorporating eMBB, URLLC and mMTC services of 5G NR. 
	

	
	\subsection{Energy efficiency in UAV-assisted cellular networks}
	The efficient utilization of the limited onboard energy of a UAV to serve cellular users is a significant challenge. Many energy-efficient solutions for UAV communication networks have been proposed to address this problem \cite{maggreen,ruieerelay,ee,alzenad20173,zeng2017energy,yang2019energy}. For instance, the authors in \cite{maggreen} developed a noncooperative game to optimize the beaconing periods among competing drones. In this way, the energy consumption in individual drones is optimized in a distributed manner. The authors in \cite{ruieerelay} proposed a spectrum and energy-efficient scheme for UAV-enabled relay network, in which UAV path is optimized by allocating the communication time slots between source and destination nodes. In \cite{ee}, the authors proposed an energy-aware power allocation scheme in the UAV-assisted edge networks while utilizing the internet of vehicles for the computation offloading. The work in \cite{alzenad20173} developed a UAV placement scheme to maximize the served users while consuming the minimum transmit power. The work in \cite{zeng2017energy} designed a UAV communication scheme to optimize the UAV trajectory while minimizing the energy consumption during the UAV flight. The work in \cite{yang2019energy} studied the joint optimization problem of user association, power allocation, and UAV deployment in UAV-enabled wireless networks.

	In order to improve energy efficiency in UAV-assisted wireless networks, it is possible to harness the possibility of \emph{energy harvesting} (particularly using renewable sources such as solar energy), a technical challenge that has not been considered in the aforementioned works \cite{maggreen,ruieerelay,ee,alzenad20173,zeng2017energy,yang2019energy}.
	By employing energy harvesting, the same UAVs can be used to serve the cellular users for a comparatively longer duration to increase the air time of UAV. In this regard, there exist only very few studies \cite{harvestalsharoaUAV,long2018energy,sekander2019performance,yang2018outage,vincent2018prospects} that discussed UAVs with energy-harvesting capabilities for wireless networking. In \cite{harvestalsharoaUAV}, the authors studied the problem of energy management in HetNets that integrate solar-powered drones. In particular, the approach in \cite{harvestalsharoaUAV} is used to optimize the trips of the drones at specific locations so as to minimize the total energy consumption in the network. The authors in \cite{long2018energy} addressed the problem of energy limitations in drones by developing a wireless power recharging system which is on renewable energy harvesting. In \cite{sekander2019performance}, a comprehensive analysis of the energy harvesting was performed for UAVs from the solar and wind sources. The problem of signal-to-noise-ratio outage minimization for the ground users was formulated to optimize the transmit power and flight duration of UAVs. In \cite{yang2018outage}, the performance analysis for the outage probability in the RF energy harvesting based UAV relaying systems was performed. However, since all the aforementioned works assume energy harvesting in UAV networks that are deployed to assist traditional wireless networks, the  study of UAV-enabled networks to satisfy the requirements of heterogeneous user traffic remains largely overlooked. In practice, it is imperative to consider the coexistence between UAV networks with an underlaid next-generation cellular infrastructure incorporating eMBB, URLLC and mMTC services.

	\subsection{Contributions}
	The main contribution of this paper is a novel energy-efficient user association and power allocation framework for optimizing energy-constrained UAV-assisted cellular networks and meet 5G NR performance requirements. To accomplish this goal, we first formulate a joint user association and power allocation problem to achieve the maximum rate for enhanced mobile broadband (eMBB) users in the network under the constraints of URLLC quality-of-service (QoS) requirements and limited power of BSs. Using the framework of \emph{ruin theory} \cite{finiteruin}, we solve the user association problem by modeling the surplus power in every UAV BS (UBS) \cite{davis2005insurance}. Here, surplus power is defined as the total residual power for a given UAV at any time instant. In particular, we determine the so-called \emph{probability of ruin} of the surplus power in each UBS and, then, we use it for user association with UBSs. 
	The adoption of ruin theory to model the surplus power in the UAVs helps to efficiently capture two opposing power flows that include: (a) the periodic power harvested in the form of regular premiums, and (b) the power allocated to the associated cellular users in the form of claims. After obtaining the energy-efficient association of the cellular users with a particular BS based on the probability of ruin, the power allocation to the associated users is performed under 5G NR constraints.

	In summary, our key contributions include the following:
	\begin{itemize}
		\item To maximize the data rate of eMBB users while satisfying the latency and reliability constraints of URLLC users, we formulate a joint energy-efficient, user association, and power allocation problem. The formulated problem is difficult to solve because the problem is \emph{mixed integer non-linear programming (MINLP)}. Therefore, we propose a novel ruin-based user association and water-filling algorithm for the power allocation to the associated eMBB users.  
		\item First, we model the surplus power in a UBS as function of three components: (a) the initial power in the UAVs at the time of dispatch, (b) the regular harvested power was obtained in the form of premiums from the solar panels mounted on the UAVs, and (c) the transmission power from UAVs to the cellular users. From the surplus process, we determine the probability of ruin which is further used for the user association.   
		\item After user association, we develop a new power allocation algorithm to allocate downlink power to the associated cellular users. Power allocation to the URLLC users is initially performed under a reliability constraint and, then, the water-filling scheme is used to allocate power to eMBB users.   
		\item Simulation results demonstrate that UAV-assisted cellular networks perform well to satisfy the stringent 5G NR requirements. Moreover, we show that the proposed ruin-based user association serves up to 58\% more number of users during a UAV flight compared to a benchmark SINR-based scheme.   
	\end{itemize}

	To the best of our knowledge, this is the first study that applies ruin theory to model the surplus power in UBSs for optimizing user association in UAV-assisted cellular networks. 
	
	The rest of this paper is organized as follows. Section \ref{secSys} presents the system model and problem formulation. In Section \ref{secSol}, we develop a ruin theoretic framework to model the surplus UAV power which is used for the user association. In Section \ref{secRA}, the proposed water-filling based power allocation algorithm is developed; and numerical evaluation of the proposed framework are discussed in Section \ref{secSim}. Conclusions are drawn in Section \ref{secCon}.

	\begin{figure}[t]
		\centering			
		\includegraphics[width=0.5\linewidth]{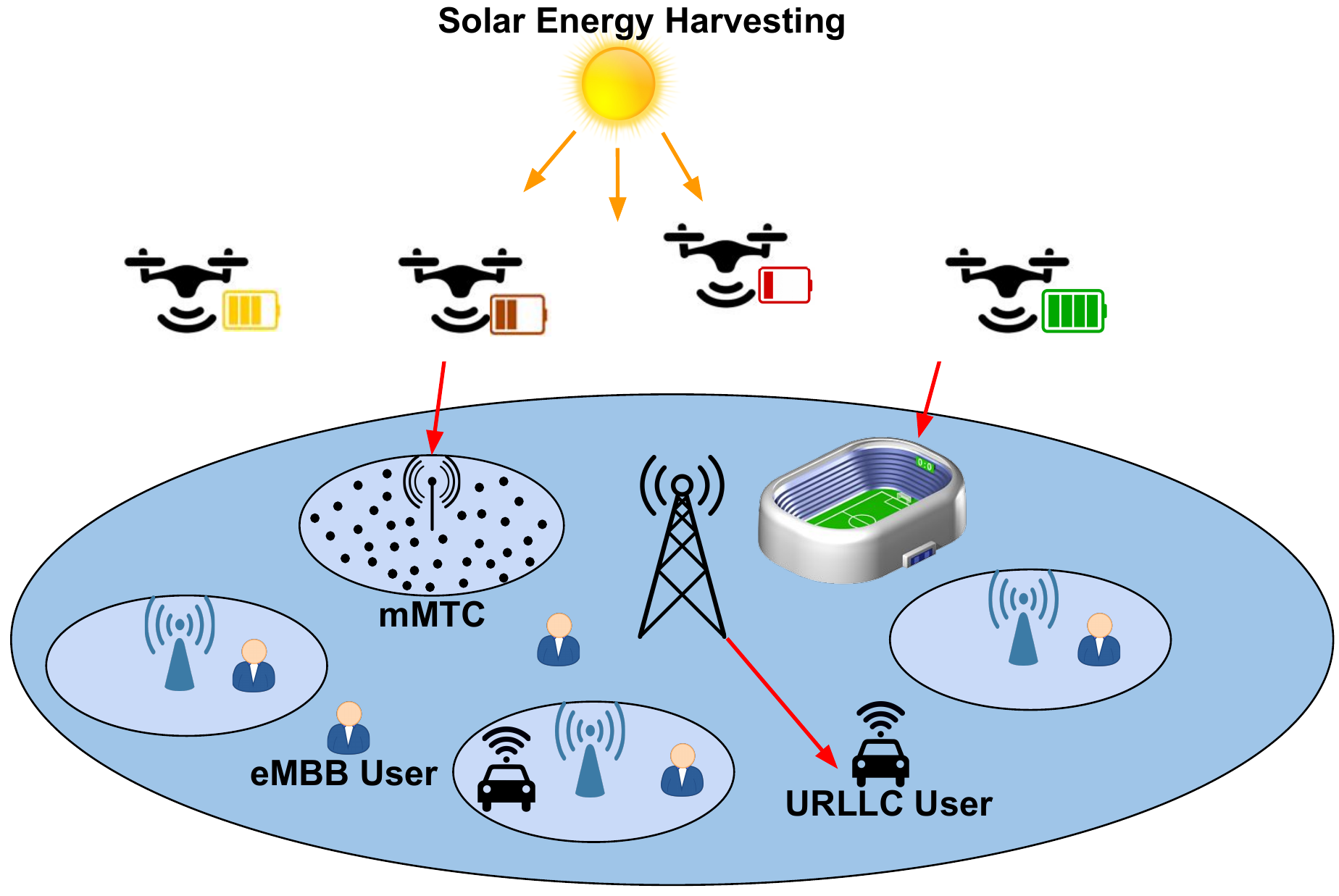}
		\caption{System model of UAV-assisted cellular network.}
		\label{sys}
	\end{figure}
	
	\section{System Model and Problem Formulation }
	\label{secSys}
	
	We consider a UAV-assisted cellular network that comprises of a single macro BS (MBS), a set $\mathcal{S}$ of $S$ SBSs, and a set $\mathcal{U}$ of $U$ UBSs for the downlink communication.  These BSs can be compositely denoted by $\mathcal{J} = \{0\} \cup \mathcal{S} \cup \mathcal{U}$, where $\{0\}$ denotes the index of the MBS. Every UAV has a limited power storage capacity to store $\rho_0$ power at the start of a flight. Moreover, there is a solar \emph{energy-harvesting module} mounted on every UAV. The UAVs are uniformly deployed above the ground cellular network at the corresponding locations denoted by $\boldsymbol{r}_u = (x_u, y_u, h_u)$, for every UAV $u \in \mathcal{U}$. We consider the following three types of users in 5G NR environment: (a) a set $\mathcal{K}_e$ of $K_e$ eMBB users, (b) a set $\mathcal{K}_u$ of $K_u$ URLLC users, and (c) a set $\mathcal{K}_m$ of $K_m$ mMTC users. The data obtained from all the IoT nodes of mMTC network is multiplexed in a single frame. The data demand from this single frame is equal to one regular eMBB user demand. Therefore, we remark that there is only one mMTC user denoted by index $\{0\}$ in the network. All these cellular users are compositely denoted by $\mathcal{K} = \{0\} \cup \mathcal{K}_e \cup \mathcal{K}_u$, where $\{0\}$ denotes the index of the mMTC user.  
	
	We study the network for a short duration of time, and the network topology is considered static during this period. For the sake of exposition, we consider that the UAVs are deployed at key locations $\boldsymbol{r}_u$ in which the cellular demand is highest. This assumption is aligned with the main use-case of UAVs as BSs that service hotspot areas for a temporary duration. We assume that determining the optimal locations follows known algorithms such as in \cite{mozaffari2016efficient}.

	\subsection{Path-loss Model}
	The signal-to-interference-plus-noise ratio (SINR), which is denoted by $\gamma_{ij}$ from the BS $j$ to the cellular user $k$ is given by the following:
	
	\begin{equation}
	\gamma_{jk} = \frac{P_{jk}h_{jk}}{\sum\limits_{j' \in \mathcal{J} \setminus \{0,j\} }P_{j'k}h_{j'k} + \sigma^2},
	\end{equation}
	where $P_{jk}$ denotes the downlink transmission power from BS $j \in \mathcal{J}$ to the user $k \in \mathcal{K}$, $h_{jk}$ denotes the channel gain, and $\sigma^2$ denotes the Gaussian noise power. Due to the different propagation environments, the channel gain of the UBSs differs from the ground SBSs and the MBS. In particular, the channel gain is given by: $h_{jk} = 10^{-\delta_{jk}/10}$ and is a function of the corresponding pathloss $\delta_{jk}$. Free space path-loss for LOS link between UAVs and ground users is considered which is expressed as $\delta_{jk} = 20\log\left( \frac{4\pi d_{jk}f}{c}\right)$, where $c$ and $f$ denote the speed of light and channel frequency, respectively.  $\delta_{jk}$, for the UBSs $j \in \mathcal{U}$ and other BSs $j \in \{ \{0\} \cup  \mathcal{S}\}$ is simplified and compositely given by the following \cite{3gpp.36.331,saadicc}: 
	\begin{equation}
	\delta_{jk} =
	\begin{cases}
	20 \log (d_{jk} f) - 147.55, &\text{for $j \in \mathcal{U}$},\\
	15.3 + 37.6 \log (d_{jk}), & \text{for $j \in \{ \{0\} \cup  \mathcal{S}\}$},
	\end{cases}
	\end{equation}
	where $d_{jk}$ denotes the distance between BS $j$ and cellular user $k$. Note that, $\delta_{jk}$ for $j \in \mathcal{U}$ is modeled as free-space path-loss (FSPL) for unobstructed LoS link between UAV $u$ and ground user $k$. The achievable rate of a cellular user $k$ associated with the BS $j$ is given by the following:
	
	\begin{equation}
	R_{jk} = x_{jk}\omega_{jk} \log\left( 1 + \gamma_{jk}\right),
	\label{eq:rate}
	\end{equation} 
	where $\omega_{jk}$ denotes the bandwidth allocated to the communication link between the BS $j \in \mathcal{J}$ and cellular user $k \in \mathcal{K}$, and $x_{jk}$ denotes the association variable given as: 
	
	\begin{equation*}
	x_{jk} =
	\begin{cases}
	1, &\text{if user $k \in \mathcal{K}$ is associated with BS $j \in \mathcal{J}$},\\
	0, & \text{otherwise}.
	\end{cases}
	\end{equation*}
	
	Each BS $j \in \mathcal{J}$ is assigned a portion of the licensed bandwidth denoted by $W_j$. This bandwidth $W_j$ is further divided equally among the various cellular users associated with BS $j$. The bandwidth $\omega_{jk} =  \frac{W_j}{ \sum\limits_{k \in \mathcal{K}} x_{jk}}$ is allocated to the communication link between the BS $j \in \mathcal{J}$ and cellular user $k \in \mathcal{K}$, where $\sum\nolimits_{k \in \mathcal{K}} x_{jk}$ denotes the total number of associated users with the BS $j$.
	

	
	\subsection{5G NR Traffic Classification Model:}
	
\begin{figure}[t]
		\centering			
		\includegraphics[width=0.5\linewidth]{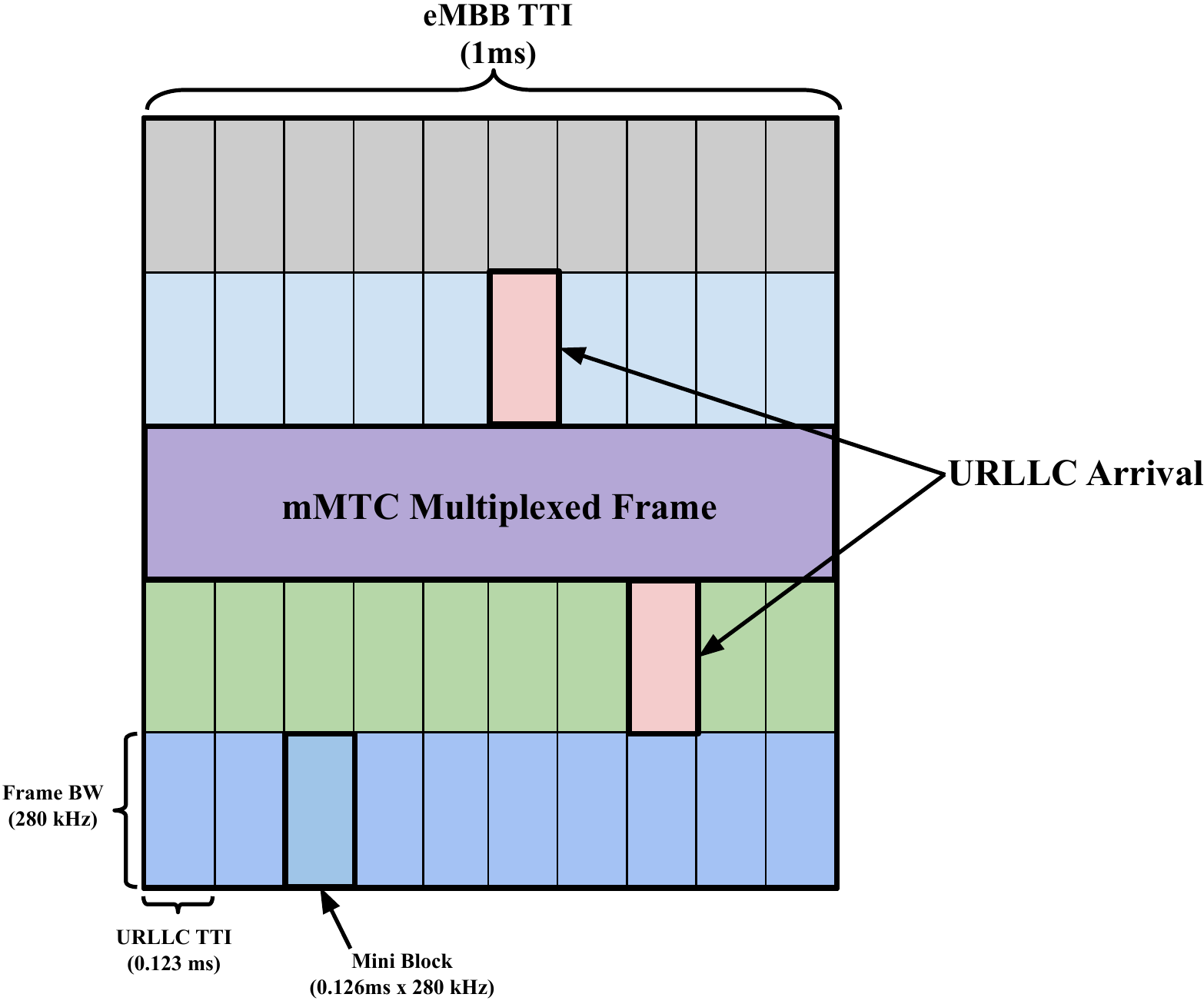}
		\caption{The frame structure of 5G NR traffic.}
		\label{5gFrame}
	\end{figure}

	As previously explained, the set of cellular users $\mathcal{K}$ in the network can be classified into two main categories based on the 5G NR traffic classification as $\mathcal{K} = \mathcal{K}_u \cup \mathcal{K}_e$ while $\mathcal{K}_u \cap \mathcal{K}_e = \emptyset $, where $\mathcal{K}_u$ denotes the set of URLLC users and $\mathcal{K}_e$ denotes the set of eMBB users. This means that any user $k$ can be classified into one of these two network user groups $\mathcal{K}_u$ or $\mathcal{K}_e$. The data obtained from all the IoT nodes of mMTC network is multiplexed in a single frame. Thus, we consider the uplink communication for the mMTC traffic which is multiplexed into a single frame and is equivalent to a regular eMBB user demand. Additionally, we consider the downlink communication for the URLLC and eMBB traffic. It can be observed from Fig. \ref{5gFrame} that the arrival of URLLC traffic reduces the achievable rate of the eMBB users because eMBB traffic could be stopped during URLLC transmission. The achievable eMBB rate after serving $\sum\nolimits_{k^{'} \in \mathcal{K}_u} x_{jk^{'}}$ URLLC requests is given as follows:  
	
	\begin{equation}
	D_{jk}\left(x_{jk} , P_{jk} \right) = \left(T-t \sum\limits_{k^{'} \in \mathcal{K}_u} x_{jk^{'}}\right) R_{jk}, \quad \forall k \in \mathcal{K}_e,
	\label{data1}
	\end{equation}
	where $T$ and $t$ denote the eMBB and URLLC TTIs respectively as shown in Fig. \ref{5gFrame}. $D_{jk}$ represents the amount of data which can be communicated from BS $j$ to the eMBB user $k$ during time $T$ while simultaneously serving $\sum\nolimits_{k^{'} \in \mathcal{K}_u} x_{jk^{'}}$ URLLC users.

	\subsection{Problem Formulation}
	
	We can now formulate an optimization problem to maximize the network rate for eMBB users subject to the URLLC reliability and latency constrains, the power level constraints of each BS $j \in \mathcal{J}$, and the unique association of the cellular users with each BS, as follows: 
	
	\begin{subequations}\label{eq:obj1}
		\begin{align}
		\max_{\boldsymbol{x}, \boldsymbol{P}} \tag{\ref{eq:obj1}} \quad 
		& \sum\limits_{j \in \mathcal{J}} \sum\limits_{k \in \mathcal{K}} D_{jk}\left(x_{jk} , P_{jk} \right) ,  \\
		&\label{eq:p1const1} \sum\limits_{k \in \mathcal{K}} P_{jk}\leq \rho_j , \quad \forall j \in \mathcal{J} , \\
		&\label{eq:p1const2} \textrm{Pr} \left( \gamma_{jk^{'}} \geq \zeta \right) \geq (1-\epsilon) , \quad \forall j \in \mathcal{J}, \forall k^{'} \in \mathcal{K}_u, \\	
		&\label{eq:p1const2_1} \sum\limits_{k^{'} \in \mathcal{K}_u} x_{jk^{'}} = \lambda_u , \quad \forall j \in \mathcal{J}, \\	
		&\label{eq:p1const3} \sum\limits_{j \in \mathcal{J}}  x_{jk} =1 , \quad \forall k \in \mathcal{K}, \\
		&\label{eq:p1const4} 0 \leq P_{jk} \leq p_{\textrm{max}} , \quad \forall j \in \mathcal{J}, k \in \mathcal{K}, \\
		&\label{eq:p1const5} x_{jk} \in \{0,1\} , \quad \forall j \in \mathcal{J}, k \in \mathcal{K}. 
		\end{align}
	\end{subequations}
	The objective is to maximize the eMBB users' sum-rate in the network. Based on the maximum power level $\rho_j$ of each BS $j$, the constraint in (\ref{eq:p1const1}) limits the total power allocation to all the associated users. (\ref{eq:p1const2}) ensures ultra-reliability by maintaining a sufficient SINR level for the URLLC users above the threshold $\zeta$ with $1-\epsilon$ confidence level. (\ref{eq:p1const2_1}) ensures low latency for URLLC users by strictly scheduling the arrived $\lambda_u$ URLLC requests in the same slot.   
	(\ref{eq:p1const3}) ensures the unique association of eMBB user $k$ with a single BS $j$. (\ref{eq:p1const4}) and (\ref{eq:p1const5}) are the bounds for the decision variables, where $p_{\textmd{max}}$ denotes the maximum power level that can be allocated for a user. Note that, the upper bound $p_{\textmd{max}}$ is helpful in reducing the interference from neighboring BSs. The following is the power level, $\rho_j$, for each BS $j \in \mathcal{J}$: 
	
	\begin{equation}
	\rho_{j} =
	\begin{cases}
	\rho_j(t), & \text{for $j \in  \mathcal{U}$},\\
	P_0, & \text{for $j = 0$}, \\
	P_1, & \text{for $j \in \mathcal{S}$}.
	\end{cases}
	\label{eq:power}
	\end{equation}
	
	\subsection{UAV Energy Model}
	
	Every UAV $u \in \mathcal{U}$ has a certain power level $\rho_u(t)$ at time instant $t$, which comprises the initial power, $\rho_0$, stored in the UAV and power $\rho$ that is harvested within a unit time. The power allocation to the set of cellular users $\mathcal{K}_u \subset K$ associated with UAV $u$ is denoted by $\sum_{k \in \mathcal{K}_u} P_{uk}$. The surplus power of UAV $u$ at time instant, $t$, is given by:
	
	\begin{equation}
	\rho_u(t) = \rho_0 + \rho t - \sum\limits_{k \in \mathcal{K}_u} P_{uk}.
	\label{p_surp}
	\end{equation} 
	
	We now observe that this surplus power has bidirectional flows of power. The positive power flow is in the form of periodic harvested power and the negative power flow is in the form of power consumption on the downlink communication from UAVs to cellular users. However, conventional stochastic optimization cannot properly capture the dynamics of these opposing powers in  UAVs. To address this challenge, we use the framework of ruin theory and model the surplus UAV power and the probability of ruin of UAV. The probability of ruin of UAV represents the vulnerability of draining out the power of a UAV. In particular, it is used to efficiently utilize the available energy of UAVs while performing user association with each UAV.
	
	\section{USER ASSOCIATION AND POWER ALLOCATION: A RUIN-BASED SOLUTION}
	\label{secSol}
	\subsection{Ruin Theory: Preliminaries}
	\label{secRuin}
	\vspace*{-0.1cm}
	In the area of actuarial science \cite{davis2005insurance}, ruin theory is applied to express an insurer's vulnerability of bankruptcy. This is performed by modeling the so-called surplus process which represents the insurer's capital at a time instant, $t$, and comprises two opposing cash flows: (a) the periodic income gained from regular insurance premiums, and (b) random claims. The insurer's vulnerability of risk is determined from the probability of ruin which essentially represents the probability of getting a negative surplus. As explained earlier, we will apply \emph{ruin theory} to formulate the surplus power in every UAV $u$ at a time instant, $t$, and is further used to compute the probability of ruin. The number of cellular users associated with the UAV is then determined based on the probability of ruin of each UAV.  
	
	\label{seSurp}
	Ruin theory is utilized to model the \emph{UAV surplus power process} which represents the random UAV power levels over time $t$. The surplus UAV power depends on three essential factors: a) launch power, b) premium power, and c) claim, that we define next. 
	\vspace*{-0.2cm} 
	\begin{mydef}
		The \emph{launch power}, which is denoted by $\rho_0$ is defined as the power stored in the UAV at the time of launch. 
	\end{mydef}	
	\begin{mydef}
		The \emph{premium power}, which is denoted by $\rho$ is defined as the constant power harvested from the renewable energy resource in a unit time. 
	\end{mydef}	
	\begin{mydef}	
		The \emph{claims} are defined as the random transmit power allocations to the cellular users associated with a UAV.   
	\end{mydef}	
	\vspace*{-0.2cm}
	The UAV surplus power, is composed of constant renewable power harvested at regular intervals, and the power dissipation on providing the communication services to the associated cellular users, i.e., $\sum\nolimits_{k \in \mathcal{K}_u} P_{uk}$. 
	
	Let, $S_u = \sum_{k \in \mathcal{K}_u} P_{uk}$ denote the compound random variable which is composed of two random variables $\mathcal{K_u}$, and $P_{uk}$, where $\mathcal{K_u}$ denotes the number of users associated with the UAV $u$ and $P_{uk}$ denotes the power allocated by the UAV $u$ to the associated cellular user $k$.  This compound variable $S_u$ is exponentially distributed with parameter $\mu$. Consequently, the surplus power given in (\ref{p_surp}) can be formally defined as follows:

	\begin{equation}
	\rho_u(t)= \rho_0 + \rho t - S_u.
	\label{surplus}
	\end{equation}

	The surplus in (\ref{surplus}), which represents the UAV power at time instant, $t$, is utilized to find the finite-time probability of ruin defined as follows \cite{finiteruin}.
	\vspace*{-0.2cm}
	\begin{mydef}
		The \emph{finite-time probability of ruin} is defined as the probability of getting a negative surplus at any time instant $s$ during a finite time $t$. The finite-time probability of ruin is mathematically given by: 
		\vspace*{-0.1cm}
		\begin{equation}
		\psi(\rho_0,t) = \normalfont \textrm{Pr} [\rho_u(s) <0 \emph{, for some $s$ as } 0 < s< t],
		\end{equation} 
		where $t$ denotes the total number of discrete time units, and $s$ ranges from $0$ to $t$. 
	\end{mydef}
	\vspace*{-0.2cm}
	In the surplus process, the depreciation in surplus is modeled using an exponential distribution of a parameter, $\mu$, which represents the distribution parameter of the claims in the surplus process. The finite-time probability of ruin, $\psi_k(\rho_0,t)$, can be obtained as follows \cite{finiteruin,aunasruin}: 
	\begin{equation}
	\psi_u(\rho_0,t) = \sum\limits_{j=1}^{t} \frac{[\mu c_j(\rho_0))]^{j-1}}{(j-1)!} e^{-\mu c_j(\rho_0)} \frac{c_1(\rho_0)}{c_j(\rho_0)},	 
	\label{eq:ruin}	
	\end{equation}  
	where $\mu$ denotes the parameter of Poisson distribution for the arrival of claims and $c_j(\rho_0) = \rho_0 + jc$, and $c_1(\rho_0) = \rho_0 +c$ represent the expressions for the accumulated power levels at time instants $j$ and $1$, respectively. In our scenario, we model the arrival of cellular users with Poisson process \cite{poisson}. $\psi_k(\rho_0,n)$ represents the probability of ruin of the UAV power at time instant, $t$, while $\rho_0$ denotes the  UAV power at time $0$.

	\subsection{Ruin-based Problem Formulation}
	
	Next, we separate (\ref{eq:p1const1}) for $j \in \mathcal{U}$ and embed it to the objective function in the form of the probability of ruin for every UAV $u$. Then we define a new optimization that is equivalent to (\ref{eq:obj1}) (the proof of equivalence is given in Appendix \ref{appProofEqui}), as follows:

	\begin{subequations}\label{eq:obj2}
		\begin{align}
		\max_{\boldsymbol{x}, \boldsymbol{P}} \tag{\ref{eq:obj2}} \quad 
		& \varsigma \sum\limits_{j \in \mathcal{J}} \sum\limits_{k \in \mathcal{K}}  D_{jk}\left(x_{jk} , P_{jk} \right) - \xi \sum\limits_{u \in \mathcal{U}}\psi_u(\rho_0,t),  \\
		&\label{eq:p2const1} \sum\limits_{k \in \mathcal{K}} P_{jk}\leq \rho_j , \quad \forall j \in \{\{0\} \cup \mathcal{S} \} , \\
		&\label{eq:p2const2} \textrm{Pr} \left( \gamma_{jk^{'}} \geq \zeta \right) \geq (1-\epsilon) , \quad \forall j \in \mathcal{J}, \forall k^{'} \in \mathcal{K}_u, \\	
		&\label{eq:p2const2_1} \sum\limits_{k^{'} \in \mathcal{K}_u} x_{jk^{'}} = \lambda_u , \quad \forall j \in \mathcal{J}, \\		
		&\label{eq:p2const3} \sum\limits_{j \in \mathcal{J}}  x_{jk} =1 , \quad \forall k \in \mathcal{K}, \\
		&\label{eq:p2const4} 0 \leq P_{jk} \leq p_{\textrm{max}} , \quad \forall j \in \mathcal{J}, k \in \mathcal{K}, \\
		&\label{eq:p2const5} x_{jk} \in \{0,1\} , \quad \forall j \in \mathcal{J}, k \in \mathcal{K}. 
		\end{align}
	\end{subequations}

	The objective is to maximize the sum-rate of the users in the network and minimize the probability of ruin for the surplus power in the UAVs, where $\varsigma$ and $\xi$ are normalizing constants. Before proceeding to the solution, we first simplify the problem by embedding (\ref{eq:p2const2_1}) in the objective function. This is done as follows:

	(\ref{eq:p2const2_1}) is an equality constraint which ensures the immediate scheduling of $\lambda_u$ number of arrived URLLC users in the same time slot $t$. This scheduling is performed by associating the URLLC user $k^{'}$ with the BS $j$ delivering best SINR $\gamma_{jk^{'}}$. The resultant eMBB rate after URLLC scheduling is expressed as follows:
	\begin{equation}
	D^{'}_{jk}\left(x_{jk} , P_{jk} \right) = \left(T-t \lambda_u \right) R_{jk}, \quad \forall k \in \mathcal{K}_e.
	\label{data2}
	\end{equation} 
	where the expression $T-t \lambda_u$ denotes the remaining duty-cycle for the eMBB communication after scheduling $\lambda_u$ number of URLLC packets. $\sum_{k^{'} \in \mathcal{K}_u} x_{jk^{'}}$ in (\ref{eq:p2const2_1}) refers to the associated URLLC users which sum up to the number of arrived URLLC users, $\lambda_u$. By replacing the expression $\sum_{k^{'} \in \mathcal{K}_u} x_{jk^{'}}$ with $\lambda_u$ in \eqref{data1}, we get \eqref{data2}. 
	
	For the optimal power allocation to the URLLC users, we state the following proposition: 
	\begin{prop}
		
		Power allocation to the URLLC users is performed to meet certain SINR threshold, $\zeta$, which ensures the URLLC reliability as given in \eqref{eq:p2const2}. The optimal power allocation to the URLLC users can be performed as follows: 
		
		\begin{equation}
		P^{*}_{jk^{'}} = \frac{ \mathcal{F}^{-1}_{\gamma_{jk^{'}}} (1-\epsilon) (1 + I )}{h_{jk}},
		\label{ualloc}
		\end{equation}
		where $\mathcal{F}^{-1}_{\gamma_{jk^{'}}} (1-\epsilon)$ denote the inverse CDF of $\gamma_{jk^{'}}$, and $I = \sum_{j' \in \mathcal{J} \setminus \{0,j\} } P_{j'k}h_{j'k} + \omega_{jk} \sigma^2$.
		\begin{proof}
			The proof of \eqref{ualloc} is given in Appendix \ref{appProofChance}.  
		\end{proof}
		
	\end{prop}
	
	Note that we analyze the network for a duration in which the power allocation from another BS $j' \in \mathcal{J} \setminus \{0,j\}$ is known. Therefore, (\ref{eq:p2const2}) becomes a convex constraint. 
	%

	After adjusting (\ref{eq:p2const2_1}) and solving the URLLC power allocation in (\ref{ualloc}), the remaining power at each BS for the eMBB users is $\rho_j - \sum_{k^{'} \in \mathcal{K}_u} P^{*}_{jk^{'}}$. The user association and power allocation problem for eMBB users is given as follows:
	\begin{subequations}\label{eq:obj4}
		\begin{align}
		\max_{\boldsymbol{x}, \boldsymbol{P}} \tag{\ref{eq:obj4}} \quad 
		& \varsigma \sum\limits_{j \in \mathcal{J}} \sum\limits_{k \in \mathcal{K}}  D^{'}_{jk}\left(x_{jk} , P_{jk} \right) - \xi \sum\limits_{u \in \mathcal{U}}\psi_u(\rho_0,t),  \\
		&\label{eq:p4const1} \sum\limits_{k \in \mathcal{K}_e} P_{jk}\leq \rho_j - \sum\limits_{k^{'} \in \mathcal{K}_u} P^{*}_{jk^{'}} , \quad \forall j \in \{\mathcal{J} \setminus \mathcal{U} \} , \\
		&\label{eq:p4const3} \sum\limits_{j \in \mathcal{J}}  x_{jk} =1 , \quad \forall k \in \mathcal{K}, \\
		&\label{eq:p4const4} 0 \leq P_{jk} \leq p_{\textrm{max}} , \quad \forall j \in \mathcal{J}, k \in \mathcal{K}, \\
		&\label{eq:p4const5} x_{jk} \in \{0,1\} , \quad \forall j \in \mathcal{J}, k \in \mathcal{K}. 
		\end{align}
	\end{subequations}
	
	It can be observed that problem (\ref{eq:obj2}) is an \emph{MINLP}, which is difficult  to solve using exhaustive search and branch and bound techniques, particularly for a large network. Furthermore, the binary variable $x_{jk}$ makes the problem combinatorial which may require exponential-complexity to solve using exhaustive search for the number of users. To avoid the difficulty, we convert the problem into simple sub-problems which are solved for each variable separately. In particular, to get a sub-optimal solution, we first fix the power allocation and solve the problem of user association by employing ruin-based heuristic approach. Then, we solve the power allocation problem using the optimal ruin-based user association solution.

	%
	
	\vspace{-0.5cm}
	
	\subsection{Ruin-based User Association with UAVs}
	\label{ruinsection}
	The first sub-problem solves $x_{jk}$ to estimate the possible number of users which can be associated with each UAV. This estimation is performed by modeling the surplus power of a UAV using the probability of ruin. After fixing $P_{jk}$, the user association sub-problem is given by:
	
	\begin{subequations}\label{eq:obj5}
		\begin{align}
		\max_{\boldsymbol{x}} \tag{\ref{eq:obj5}} \quad 
		& \varsigma \sum\limits_{j \in \mathcal{J}} \sum\limits_{k \in \mathcal{K}}  D^{'}_{jk}\left(x_{jk} , P_{jk} \right) - \xi \sum\limits_{u \in \mathcal{U}}\psi_u(\rho_0,t),  \\
		&\label{eq:p5const1} \sum\limits_{j \in \mathcal{J}}  x_{jk} =1 , \quad \forall k \in \mathcal{K}, \\	
		&\label{eq:p5const2} x_{jk} \in \{0,1\} , \quad \forall j \in \mathcal{J}, k \in \mathcal{K}. 
		\end{align}
	\end{subequations}
	

	
	For the sake of energy-efficiency, the user association $x_{jk}$, with every UAV $u$ is performed based on its probability of ruin, $\psi_u(\rho_0,t)$, obtained from (\ref{eq:ruin}). This is done by limiting the number of associated users with each UAV based on the probability of ruin. Meanwhile, when the probability of ruin of a UAV is high, fewer users are associated to that UAV and vice versa. One meaningful approach for user association is to associate the cellular user, $k$, with that BS which is providing better SINR level, $\gamma_{jk}$. This SINR-based approach is not energy-efficient and, hence, it is not suitable for the problem of power constrained user association with the UAVs. We propose an approach that combines SINR and probability of ruin-based approach in the following definition.  
	
	\begin{mydef}
		To incorporate both the SINR and the probability of ruin in the UAV user association problem, we introduce a factor, $\eta_{jk}$, defined as follows:
		
		
		\begin{equation}
		\eta_{jk} := \alpha (1-\psi_u(\rho_0,t)) \gamma_{jk},
		\label{eta}
		\end{equation}
		where the term $(1-\psi_u(\rho_0,t))$ denotes the probability of survival and $\alpha$ is a control factor. 
	\end{mydef}	
	It can be observed that the factor $\eta_{jk}$ consolidates both the SINR and the probability of ruin. Therefore, the cellular user is associated with that UBS that provides a better SINR to the cellular user and has a relatively less probability of ruin.

	\begin{algorithm}[t]
		\caption{User Association Algorithm}\label{assAlgo}
		\begin{algorithmic}[1]		
			\State \textbf{Input:} $J$, $K$, $P_{jk}$, $\rho_j$		
			\State \textbf{initialize}: $x^{*}_{jk}=0$
			\State \textbf{Step 1:}
			\State Compute $\psi_u(\rho_0,t)$ from \eqref{eq:ruin}
			\State Compute $\eta_{jk}$ from \eqref{eta}
			\For{$k = 1$ \textbf{to} $K$}  
			\State Select single BS $j$ with $\max\limits_{j \in \mathcal{J}} \eta_{jk}$
			\EndFor
			\State \textbf{Step 2:}
			\For{$j = 1$  \textbf{to} $J$}
			\State Initialize $P = \rho_j$	
			\While{$P \geq 0$}
			\State Find $\max\limits_{k \in \mathcal{K}} \gamma_{jk}$
			\State Update $x^{*}_{jk} = 1$, and $P = P - P_{jk}$ \label{alg:goto}
			\State Remove $\max\limits_{k \in \mathcal{K}} \gamma_{jk}$ from $SINR$ vector $\gamma_{jk}$
			\EndWhile
			\EndFor	
		\end{algorithmic}
	\end{algorithm}

	The user association problem \eqref{eq:obj5} is combinatorial and difficult to solve for the large network. To solve the problem, we present a ruin-based heuristic approach in Algorithm \ref{assAlgo} that results to a sub-optimal solution. The inputs are the following: the total number of BSs, $J$, the total number of network users, $K$, fixed power allocation $P_{jk}$, allocated to each user $k$, and the total power bound, $p_j$ of BS $j$. In Algorithm \ref{assAlgo}, first, the probability of ruin, $\psi_u(\rho_0,t)$ is computed, which is further used to compute $\eta_{jk}$ from \eqref{eta}. Then, every cellular user, $k$, selects the BS $j$ with the maximum value of $\eta_{jk}$. Next, the users with better SINR values are associated with the corresponding BS based on the power level $\rho_j$ of each BS. The association $x_{jk}^{*}$ and the remaining BS power $P$ is updated in step $14$. The algorithm terminates when either $P$ is zero or all the cellular users are associated. This algorithm gives a sub-optimal solution with the complexity $\mathcal{O}(N)$ for performing user association using the probability of ruin. We will show the optimality gap in the simulation results in Section \ref{secSim}, where it will be observed that the UAV with less surplus power will have a high probability of ruin. Hence, fewer users will be associated with that UAV.  
	
	\vspace{-0.5cm}
	
	\subsection{Power allocation to eMBB users}
	\label{secRA}
	
	From the previous section, we obtain the optimal associated users denoted by $x_{jk}^*$. After performing the ruin-based optimal user association, the next step is to solve the power allocation problem. Thus, the updated value of the achievable rate for the set of the associated eMBB users is given by:
	
	\begin{equation}
	R^{'}_{jk} = \left(T-t \lambda_u \right) x^{*}_{jk}\omega_{jk} \log\left( 1 + \gamma_{jk}\right).
	\end{equation} 
	
	The proof of equivalence of \eqref{eq:obj4} and \eqref{eq:obj3} is given in Appendix \ref{secproof2}.  The sub-problem for the power allocation to eMBB users is given as follows:
	
	\begin{subequations}\label{eq:obj3}
		\begin{align}
		\max_{ \boldsymbol{P}} \tag{\ref{eq:obj3}} \quad 
		& \sum\limits_{j \in \mathcal{J}} \sum\limits_{k \in \mathcal{K}_e}  R^{'}_{jk},  \\
		\text{s.t.}\quad 
		&\label{eq:p3const1} \sum\limits_{k \in \mathcal{K}_e} P_{jk}\leq \rho_j - \sum\limits_{k^{'} \in \mathcal{K}_u} P^{*}_{jk^{'}} , \quad \forall j \in \{\{0\} \cup \mathcal{S} \} , \\
		&\label{eq:p3const3} 0 \leq p_{jk} \leq p_{\textrm{max}} , \quad \forall j \in \mathcal{J}, k \in \mathcal{K}_e.
		\end{align}
	\end{subequations}
	
	\begin{algorithm}[t]
		\caption{Iterative Ruin-Based Resource Allocation Algorithm}\label{finAlgo}
		\begin{algorithmic}[1]		
			\State \textbf{Input:} $J$, $K$, $P_{jk}$, $\rho_j$, $\eta_{jk}$		
			\State Set uniform power to each BS and user pair \{j,k\} i.e.\ $P_{jk}[0] = \rho_j / K$ 
			\State Compute $\gamma_{jk}$ at each user $k$ from the BS $j$
			\State \textbf{Initialize}: $t=0$
			\State Associate and allocate power to URLLC users using (\ref{ualloc})		
			\While {$t\leq T_{\textrm{max}}$ \textbf{or} $ \epsilon^{*} \geq \epsilon_0 $  } 
			\State $x_{jk}^{*}[t] = \underset{x}{\mathrm{argmax}}\sum\limits_{j \in \mathcal{J}} \sum\limits_{k \in \mathcal{K}}  D^{'}_{jk} $
			\State Allocate power $P_{jk}^{*}$ to eMBB users using (\ref{wtfill}) 
			\State \textbf{Update}
			\State $P_{jk}[t] = P_{jk}^{*}$
			\State $ \epsilon^{*} =  P_{jk}[t]- P_{jk}[t-1] $
			\State $t = t + 1$
			\EndWhile	
		\end{algorithmic}
		\label{algo2}
	\end{algorithm}
	
\begin{figure}[t]
		\centering			
		\includegraphics[width=0.5\linewidth]{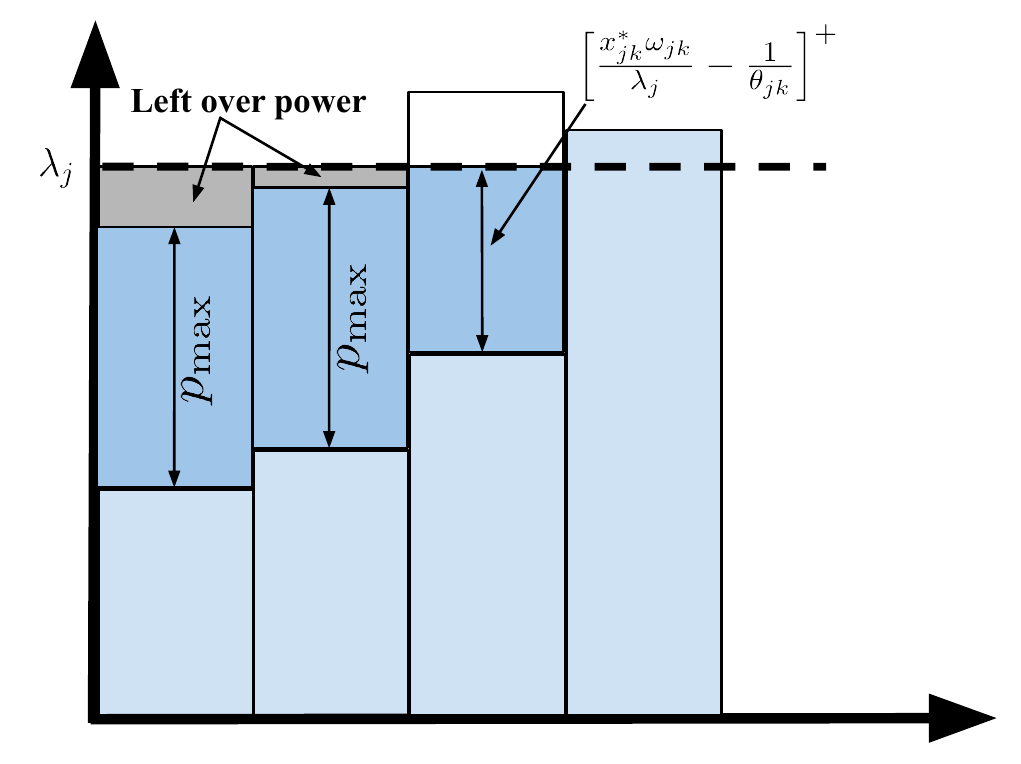}
		\caption{An illustration of the water-filling algorithm.}
		\label{wfill}
	\end{figure}

	Problem (\ref{eq:obj3}) is convex because the objective function and all of the constraints are convex. To solve this problem, we use a variant of the water-filling algorithm where the power, $P_{jk}$ allocated to a user is limited by a threshold, $p_{\textrm{max}}$, and the leftover power is allocated to the other low-gain network users as shown in Fig. \ref{wfill}. The following optimal solution is obtained:
	
	\begin{prop}
		
		The optimal power allocation $P_{jk}^{*}$ is expressed as: 
		
		\begin{equation}
		P_{jk}^{*} = \min \left\{p_{\rm{max}},
		\left[\frac{ x^{*}_{jk}\omega_{jk}}{\lambda_j}  - \frac{1}{\theta_{jk}} \right]^+ \right\} , \quad \forall j \in \mathcal{J}, k \in \mathcal{K}_e,
		\label{wtfill}
		\end{equation}
		where $\theta_{jk}$ is the channel gain for the user $k$ from BS $j$ defined as:
		\begin{equation}
		\theta_{jk} = \frac{h_{jk}}{1 + \sum\limits_{j' \in \mathcal{J} \setminus \{0,j\} }P_{j'k}h_{j'k} + \omega_{jk} \sigma^2}, \quad \forall j \in \mathcal{J}, k \in \mathcal{K}_e.
		\end{equation}
		
		We note that the maximum power that can be allocated to a user is limited by $p_{\rm{max}}$. From (\ref{wtfill}), we get $\lambda_j^*$ which denotes the optimal water level chosen such that the following condition is satisfied. 
		\begin{equation}
		\sum\nolimits_{k \in \mathcal{K}_e} P_{jk}^{*} = \rho_j - \sum\limits_{k^{'} \in \mathcal{K}_u} P^{*}_{jk^{'}}, \quad \forall j \in \mathcal{J}. 
		\label{eq:lam}
		\end{equation}
		\begin{proof}
			See in Appendix \ref{secproof1}. 
		\end{proof}
	\end{prop}
	
	The maximum power bound $p_{\textrm{max}}$ helps to serve more users by allocating the leftover power to low channel gain users. This is performed by adjusting the water level, $\lambda_j$, using (\ref{eq:lam}). This is illustrated in Fig. \ref{wfill} where the first two users are allocated maximum power, while the leftover power is allocated to the other users by accordingly adjusting the water level, $\lambda_j$. The third user is allocated power using (\ref{wtfill}) based on the available channel gain, $\theta_{jk}$, while the fourth user is not allocated any power because of a very small channel gain. 
	
	\begin{figure}[t]
	\centering			
	\includegraphics[width=0.5\linewidth]{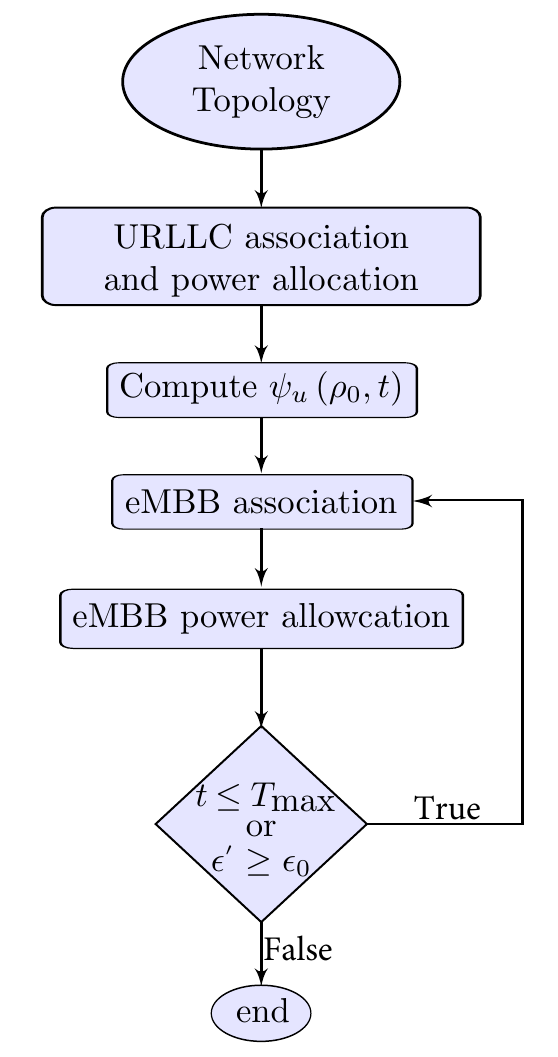}
	\caption{Systematic diagram of the iterative algorithm.}
	\label{flow}
	\end{figure}
	
	Algorithm \ref{algo2} solves \eqref{eq:obj4} for association and power allocation by iteratively solving the association and power allocation problems. First, a uniform power is allocated to all the user-BS pairs. Then, the users maximizing the network rate under maximum power bounds are selected. Then, the power allocation for the selected users is performed using the water-filling algorithm. Fig. \ref{flow} shows the systematic diagram and flowchart of the iterative algorithm for the association and power allocation. Note that this algorithm uses water-filling algorithm with the complexity $\mathcal{O}(N.M)$. The algorithm gives a sub-optimal solution and we show the optimality gap in the next section.

	\begin{table}[t]
		\centering
		\begin{tabular}{|c|c|c|c|}
			\hline
			\textbf{Parameters}   & \textbf{Values}  & \multicolumn{1}{l|}{\textbf{Parameters}} & \multicolumn{1}{l|}{\textbf{Values}} \\ \hline
			URLLC TTI          & $0.125$~ms  & eMBB TTI                     & $1$~ms                           \\ \hline
			MBS Radius            & $500$~m     & Noise ($\sigma^2$)                                & -97.5 dBm                            \\ \hline
			SBS Radius            & $100$~m     & Frequency (f)                            & $2$~GHz                         \\ \hline
			UBS Height         & $200$~m     & Bandwidth $W$                            & $50$~MHz                        \\ 
			\hline
			\multicolumn{2}{|c|}{UAV Transmit Power} & \multicolumn{2}{c|}{$0.5$~W}                              \\ 
			\hline
			\multicolumn{2}{|c|}{Path Loss (UBS)} & \multicolumn{2}{c|}{$20 \log (d_{uk} f) - 147.55$}                              \\ \hline
			\multicolumn{2}{|c|}{Path Loss (SBS and MBS)}    & \multicolumn{2}{c|}{$16.62+37.6 \log(d))$}                                        \\ \hline
		\end{tabular}
		\caption{Simulation parameters}
		\label{tab1}
	\end{table}

	\section{Simulation Results}
	\label{secSim}

	For our simulation, we consider a geographical area of $4000$~m $\times$ $4000$~m square. MBS is deployed in the center at a fixed location, whereas 10 SBSs and 5 UAVs are uniformly deployed in the area. The cellular users are randomly located in the geographical area. Statistical results are averaged over several runs of random locations of cellular users, SBSs and UAVs. Other simulation parameters are given in table \ref{tab1}. 
	
	\begin{figure}[t]
	\centering			
	\includegraphics[width=0.5\linewidth]{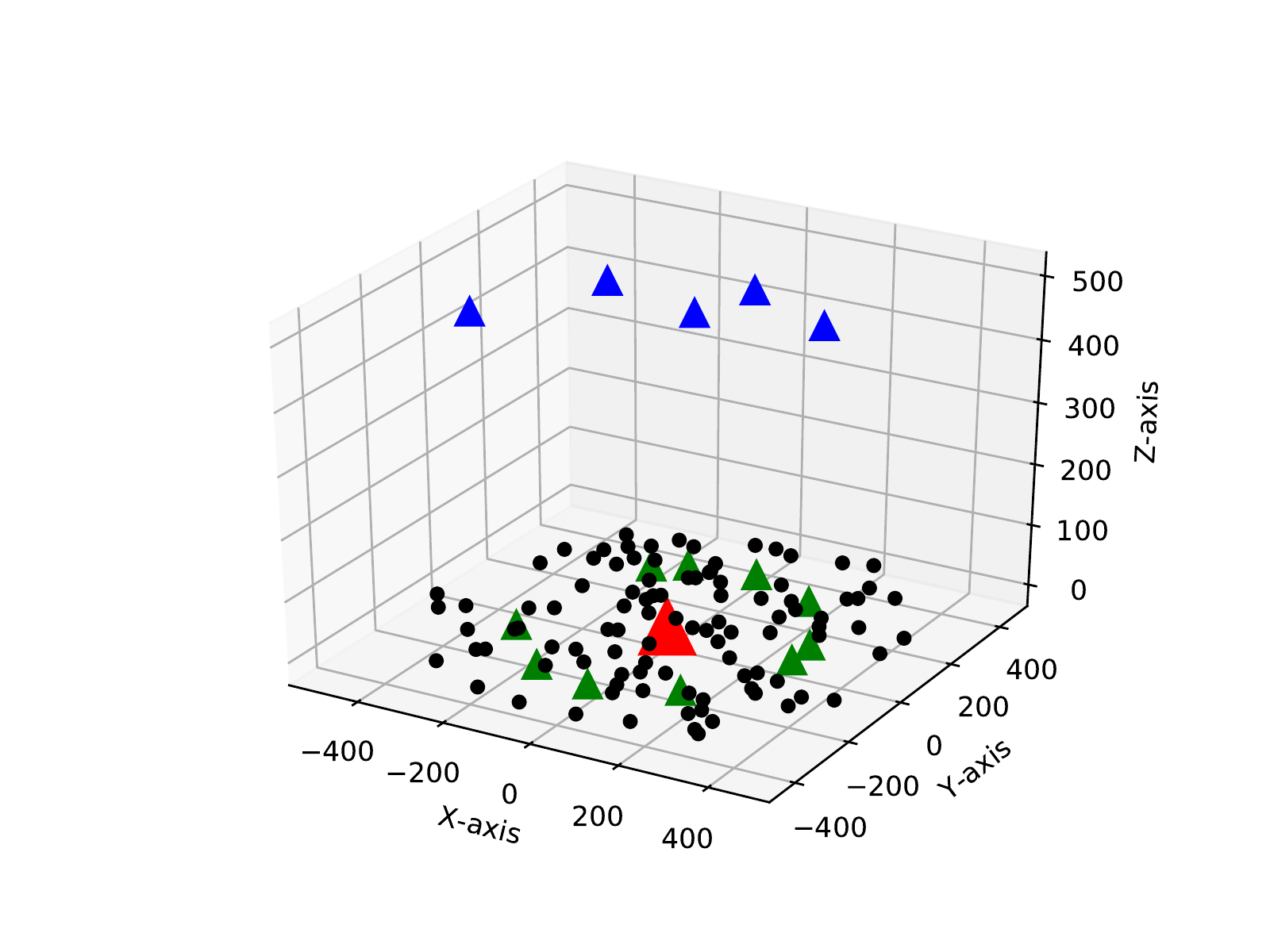}
	\caption{Network topology consisting of MBS, SBSs, UBSs and the cellular users.}
	\label{res0}
	\end{figure}
	
	Fig. \ref{res0} shows the network topology which comprises a single MBS and uniformly deployed SBSs, UBSs and the cellular users. The UBSs are deployed at a height of $200$~m above the ground level.

	\begin{figure}[t]
		\centering			
		\includegraphics[width=0.5\linewidth]{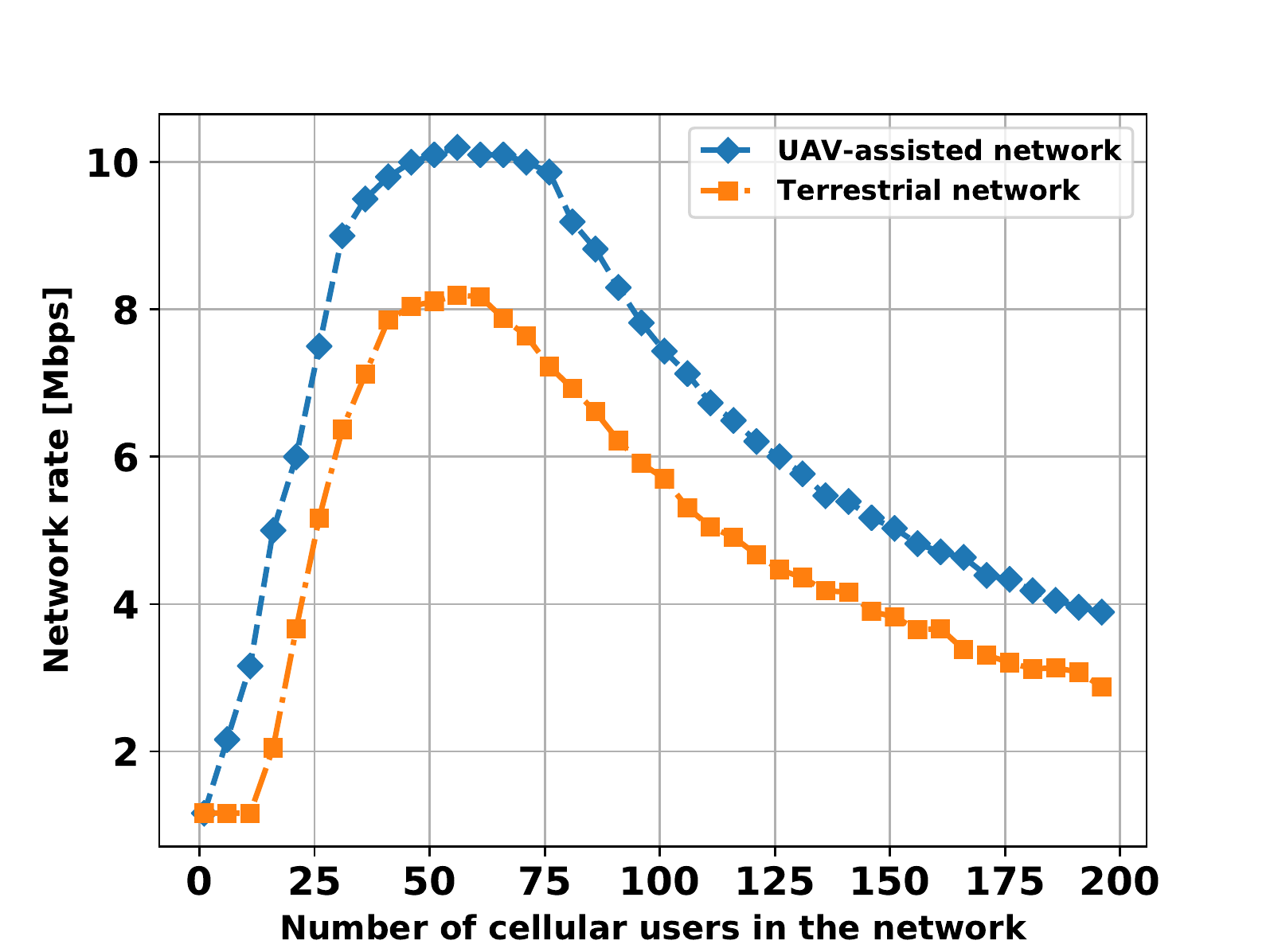}
		\caption{Network rate vs. number of cellular users in the network.}
		\label{cellRate}
	\end{figure}
	
	Fig. \ref{cellRate} demonstrates the per-user network rate as a function of the number of users. The network rate is calculated after solving the optimization problem for $5$ UAVs and $5$ SBSs. These results indicate that the UAV-assisted network achieves a better per-user rate when compared with the terrestrial network. For example, when the number of users in the network is about $75$, the UAV-assisted network achieves about $40$\% more rate when compared with the terrestrial network. This is due to the LoS communication link between UAV and ground users which deliver better SINR as compare to the non-LoS link of the terrestrial network. From Fig. \ref{cellRate}, we can also observe that the network rate reduces as the number of users in the network increases. This is due to the limited power resources at BSs which are insufficient to serve all the users in the network.     
	
	\begin{figure}[t]
		\centering			
		\includegraphics[width=0.5\linewidth]{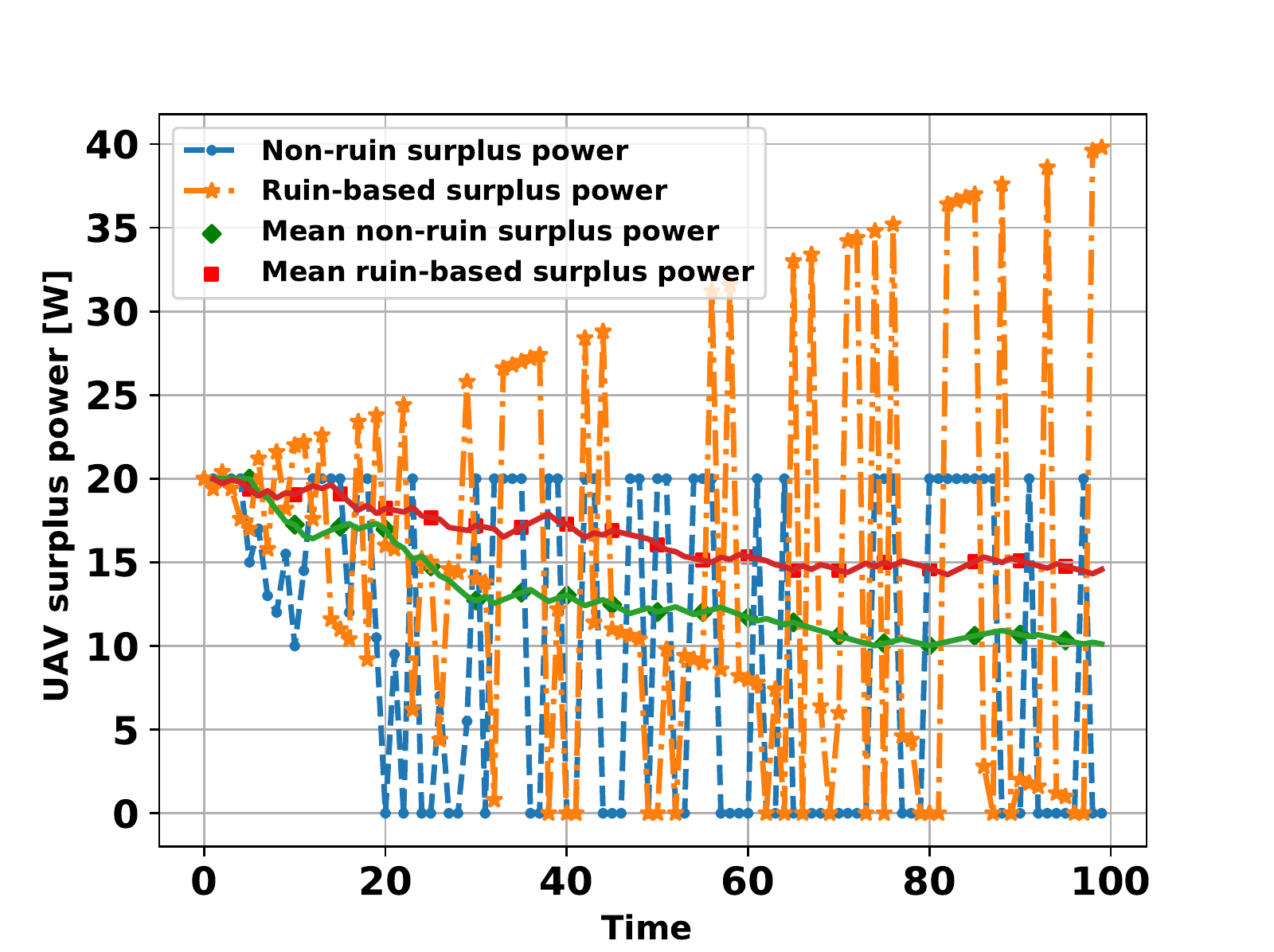}
		\caption{Total surplus power of all UAVs at different time instants.}
		\label{surpResult}
	\end{figure}
	
	Fig. \ref{surpResult} illustrates how the surplus power per UAV varies over time during a single flight of UAV. We show the power allocation to the associated users for $100$ TTIs. The optimization problem is solved by running the ruin-based association algorithm and water-filling power allocation algorithm for each TTI. The initial UAV surplus power at the time of launch $t = 0$ is set to be $100$~Watts. 
	By controlling the number of the associated users through the probability of ruin, up to 52\% higher level of the UAV surplus power is achieved when compared with the non-ruin approach. By preserving the surplus UAV power as shown by the significant differences in power drops at certain time instants, the ruin-based approach can serve more number of cellular users eventually.    
	
	\begin{figure}[t]
		\centering			
		\includegraphics[width=0.5\linewidth]{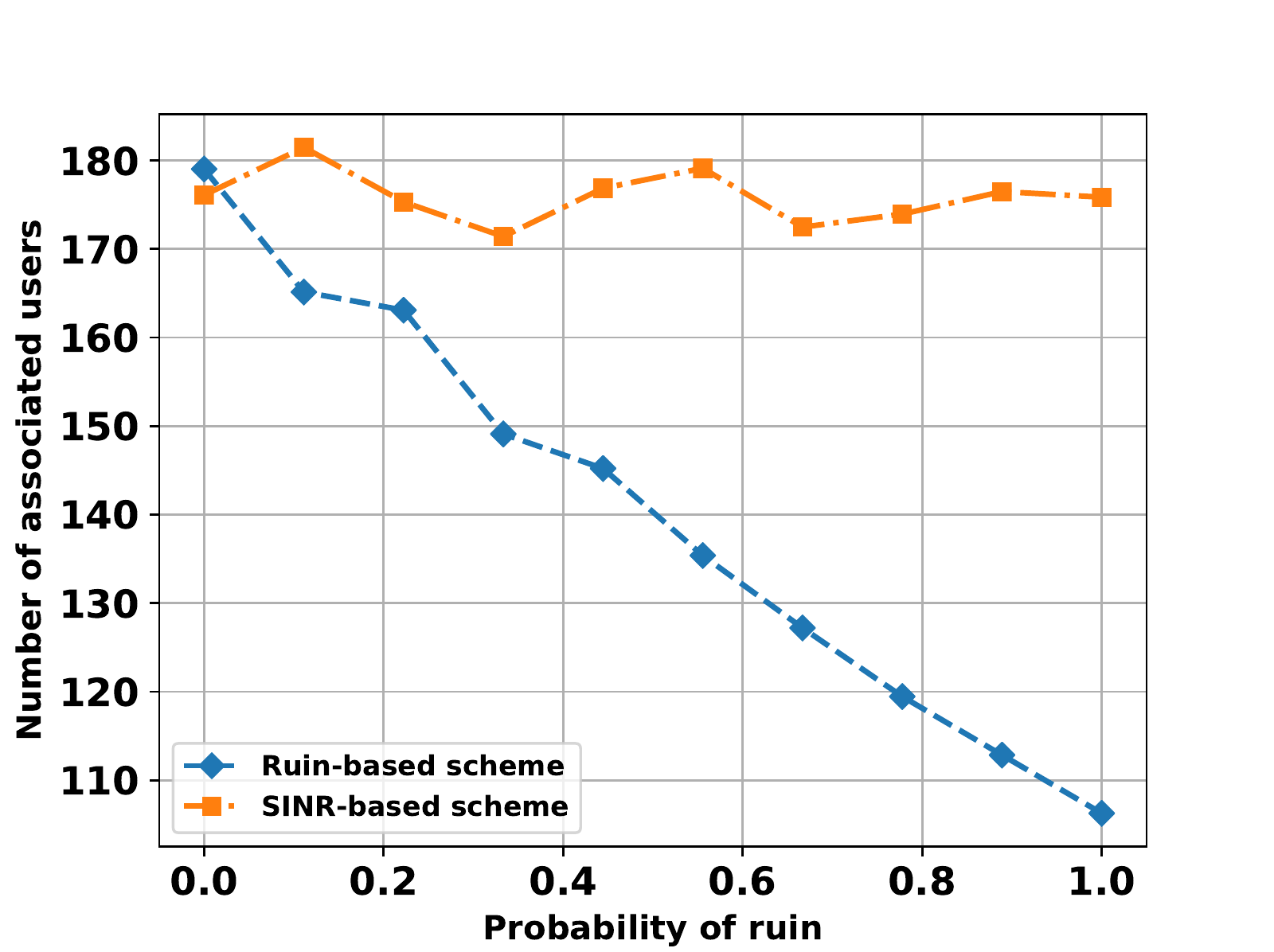}
		\caption{Number of total users associated with all UAVs vs. the probability of ruin.}
		\label{ruinResult}
	\end{figure}
	
	Fig. \ref{ruinResult} demonstrates how the number of UAV-associated cellular users varies with the probability of ruin. We compare the proposed ruin-based scheme with a baseline SINR-based scheme, which associates the cellular users with the UAVs to give the best SINR without considering the surplus power of the UAVs. By decreasing the number of associated cellular users with the UAVs, it can be observed that the proposed ruin-based scheme tries to preserve UAV power with an increase in the probability of ruin. The non-associated users are offloaded to other UAVs with less probability of ruin or to the terrestrial network.  
	
	\begin{figure}[t]
		\centering			
		\includegraphics[width=0.5\linewidth]{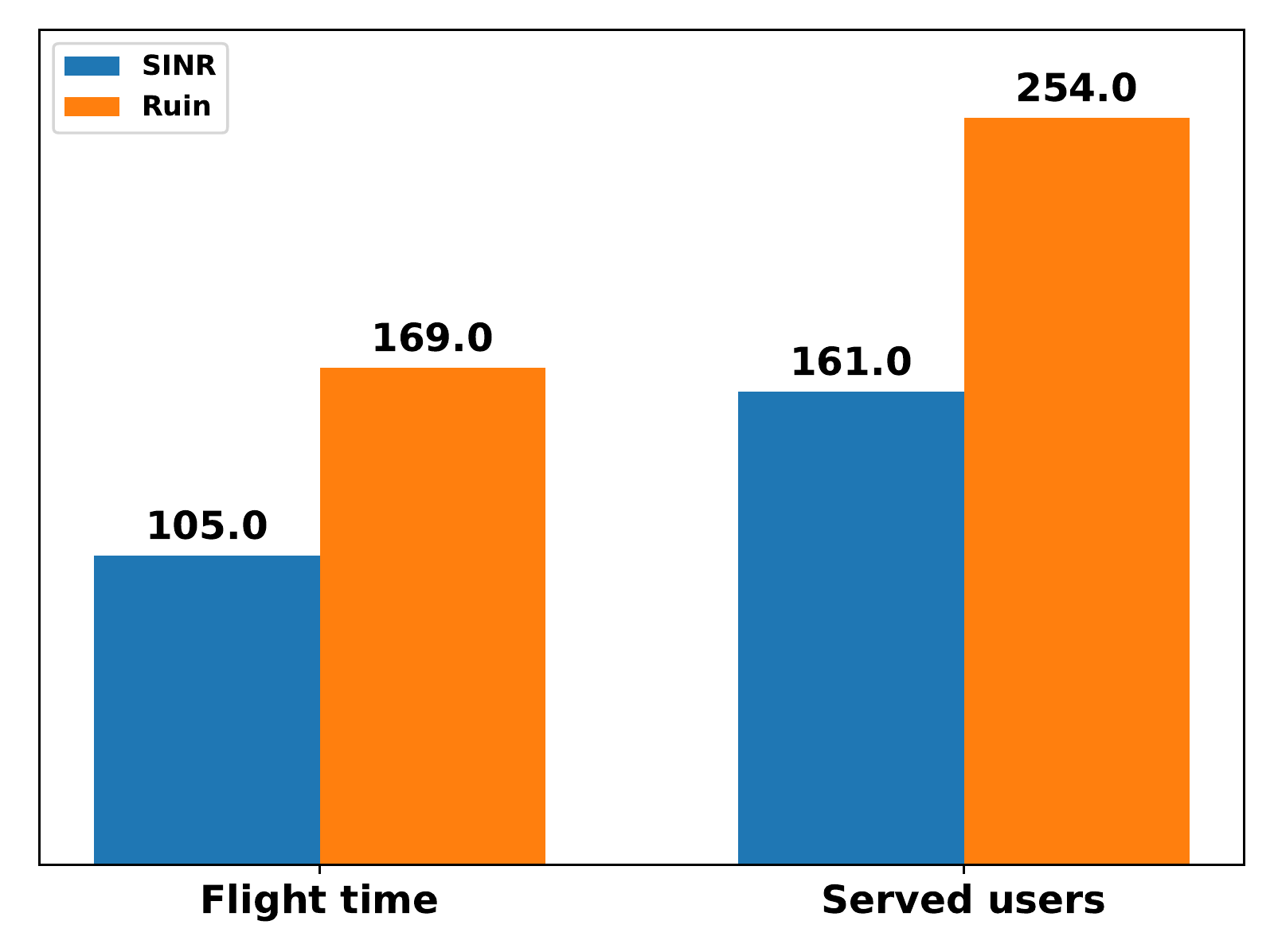}
		\caption{Comparison of ruin and SINR-based approach for UAV flight time and number of served users.}
		\label{ruinResult2}
	\end{figure}
	
	Fig. \ref{ruinResult2} shows the flight time in units of TTIs and the number of served users during a single flight of UAV for the SINR-based and ruin-based approaches. Fig. \ref{ruinResult2} demonstrates that the ruin-based approach enhances the flight time of UAV by offloading the users to terrestrial network. As a result, exploiting the energy harvested during the enhanced flight duration, more number of cellular users are served by the UAV eventually. For instance, the flight duration of UAV in the ruin-based approach is $64$~TTIs as compared to the SINR-based approach. During $64$ TTIs, additional energy is harvested which is further used to serve $93$ more users during the UAV flight.      
	
	\begin{figure}[t]
		\centering			
		\includegraphics[width=0.5\linewidth]{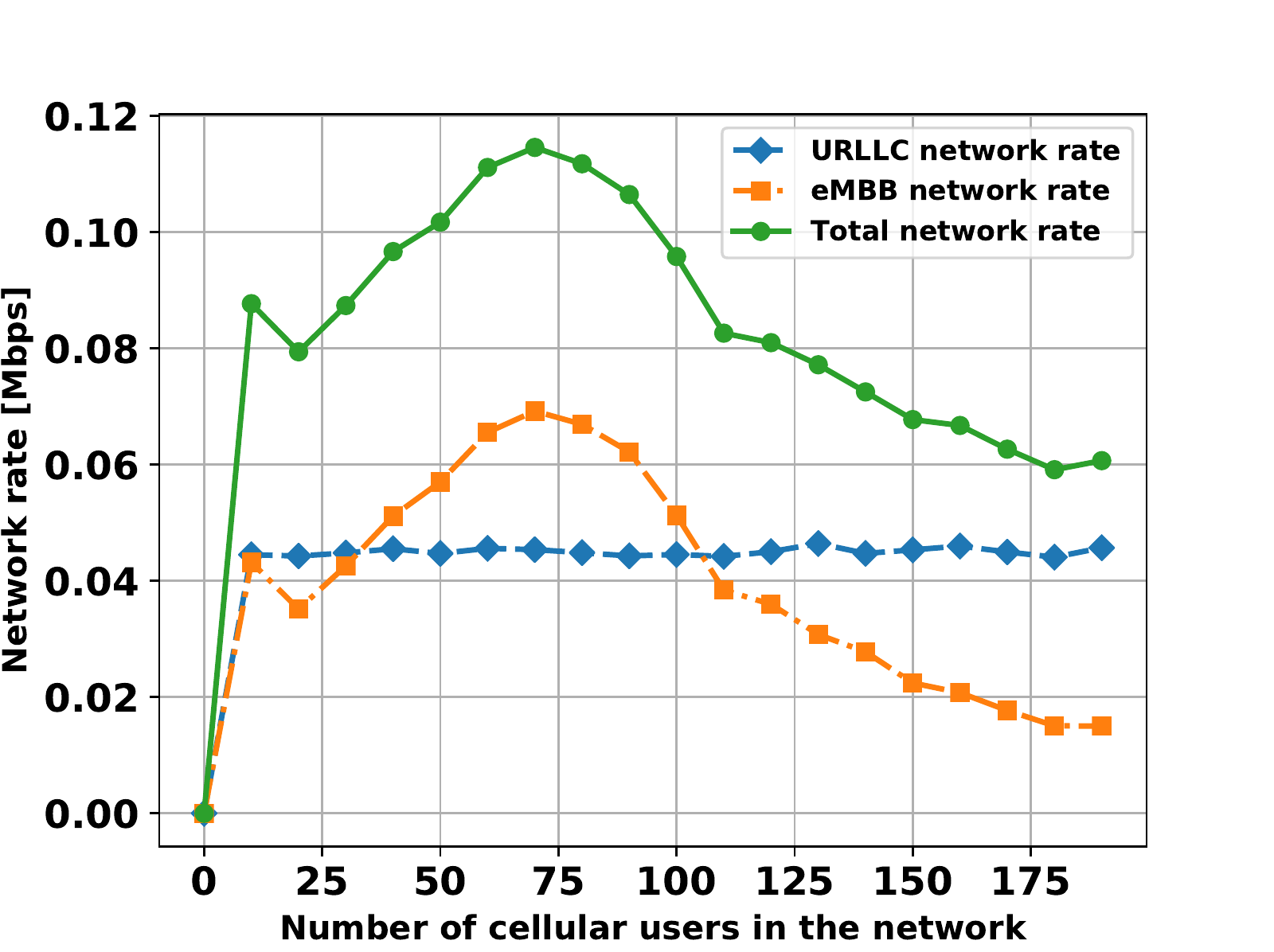}
		\caption{Network rate vs. number of cellular users in the network.}
		\label{5gRate}
	\end{figure}
	
	Fig. \ref{5gRate} shows the plot of a network for the two traffic classes of 5G NR, i.e.,\ eMBB and URLLC against the number of network users. From this figure, we can see that the rate of URLLC traffic is not affected by increasing the number of network users. Meanwhile, the eMBB rate significantly decreases as the number of users in the network increases. This is because the URLLC users are given priority over the eMBB users by allocating the resources to satisfy their latency and reliability requirements.   
	
	\begin{figure}[t]
		\centering			
		\includegraphics[width=0.5\linewidth]{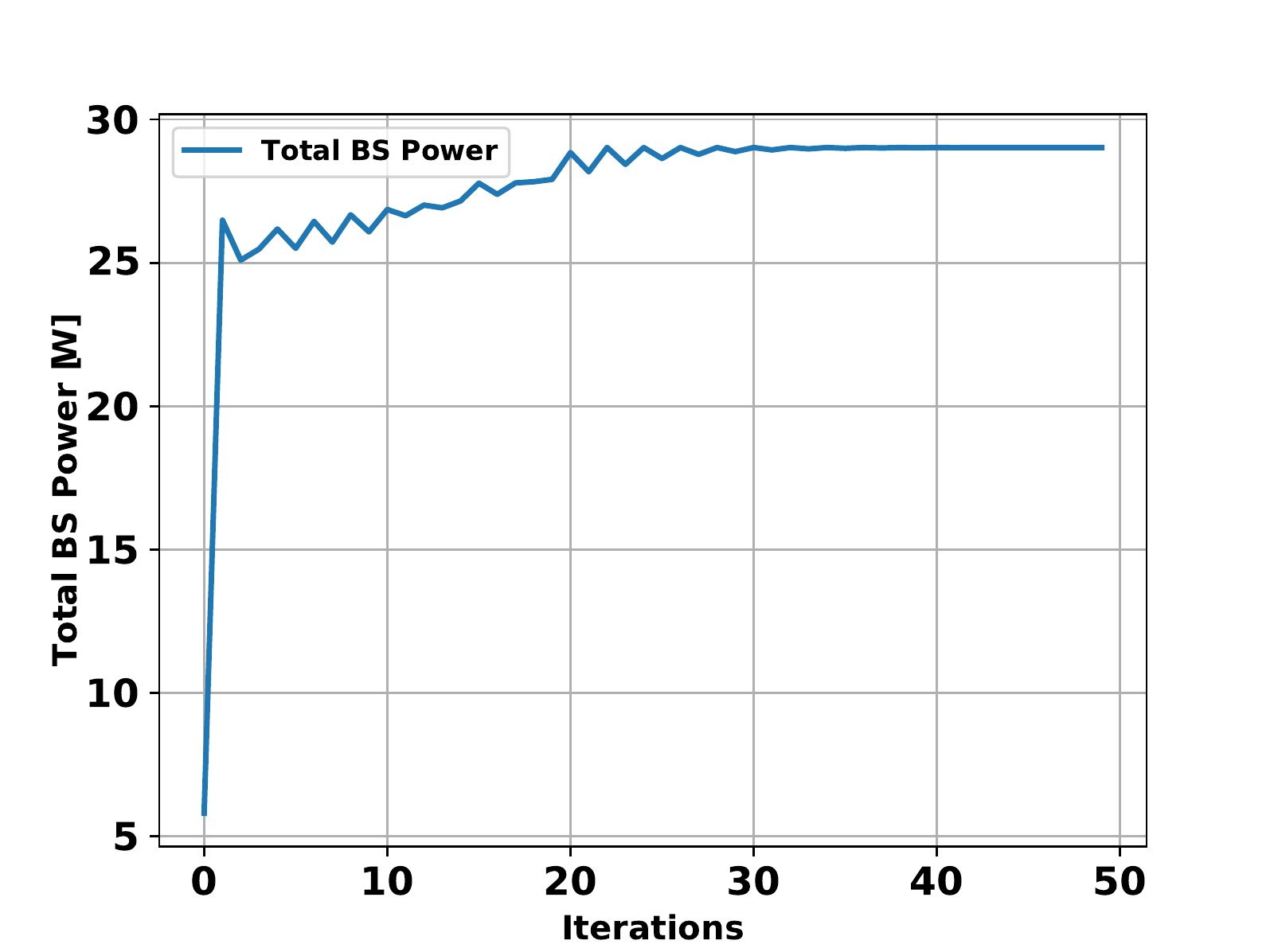}
		\caption{Plot of convergence.}
		\label{convergence}
	\end{figure}
	
	Fig. \ref{convergence} shows the convergence of power allocation algorithm against the number of iterations. The power allocation is performed to the associated set of users by the BS in an iterative manner. It can be observed that the algorithm converges after $30$ number of iterations. 
	
	\begin{figure}[t]
		\centering			
		\includegraphics[width=0.5\linewidth]{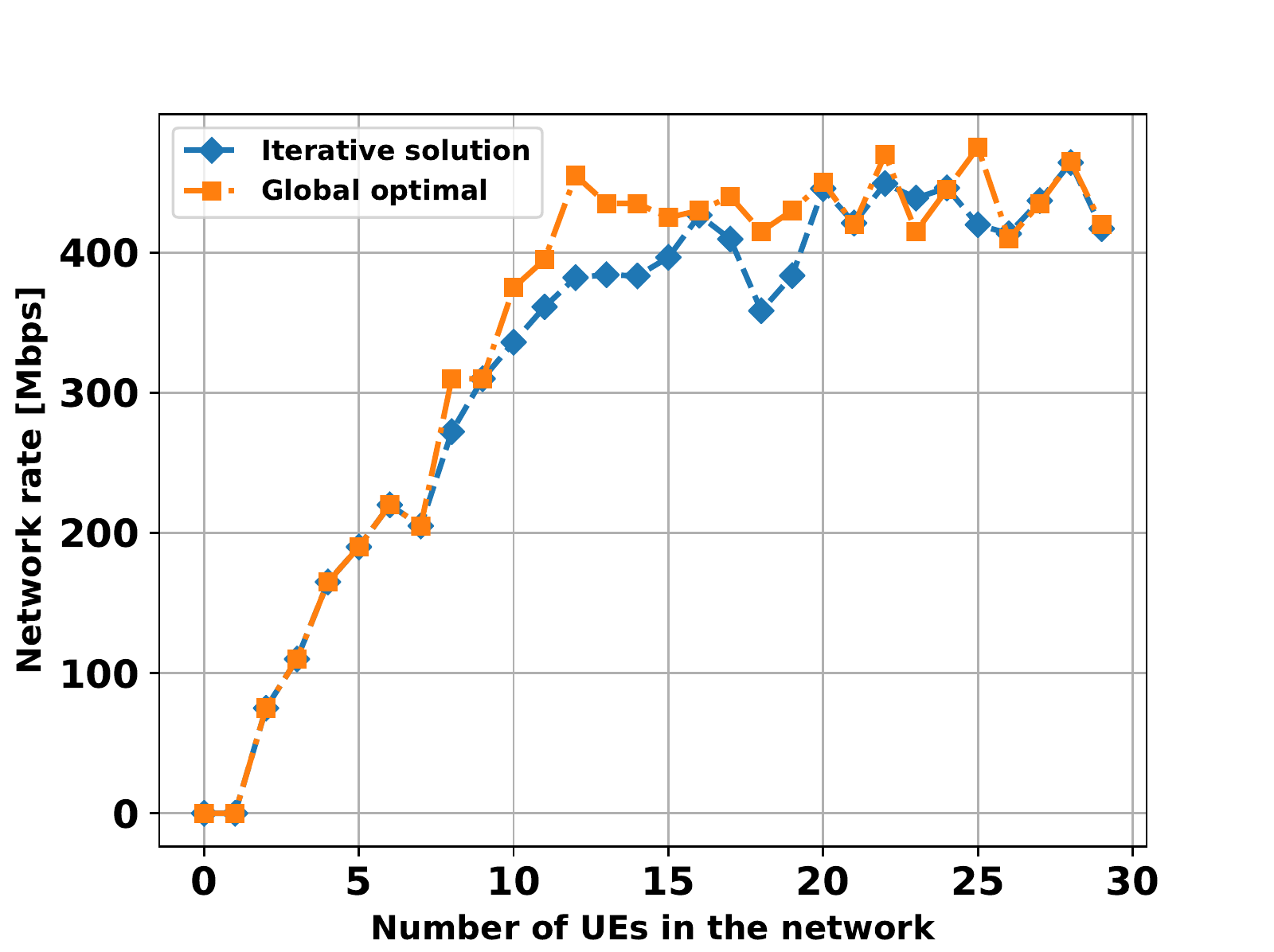}
		\caption{Network rate vs. number of cellular users in the network.}
		\label{optgap}
	\end{figure}
	
	Fig. \ref{optgap} demonstrates the optimality gap between the proposed and optimal solution. We use the Gurobi optimization tool to obtain the optimal solution of the original problem \ref{eq:obj1}. We show the optimality gap results for the small network containing up to $30$ cellular users. It can be observed that there is no gap for a small network when there are up to $7$ cellular users in the network. A very small and negligible gap is observed as the number of cellular users increases in the network. This gap is due to the upper bound of maximum number of iterations $T_{\textrm{max}}$. This upper bound restricts the iterative algorithm to the sub-optimal solution for the large network.   
	
	\begin{figure*}[t]	
		\subfigure[\label{WF1}]{\includegraphics[width=0.3\linewidth]{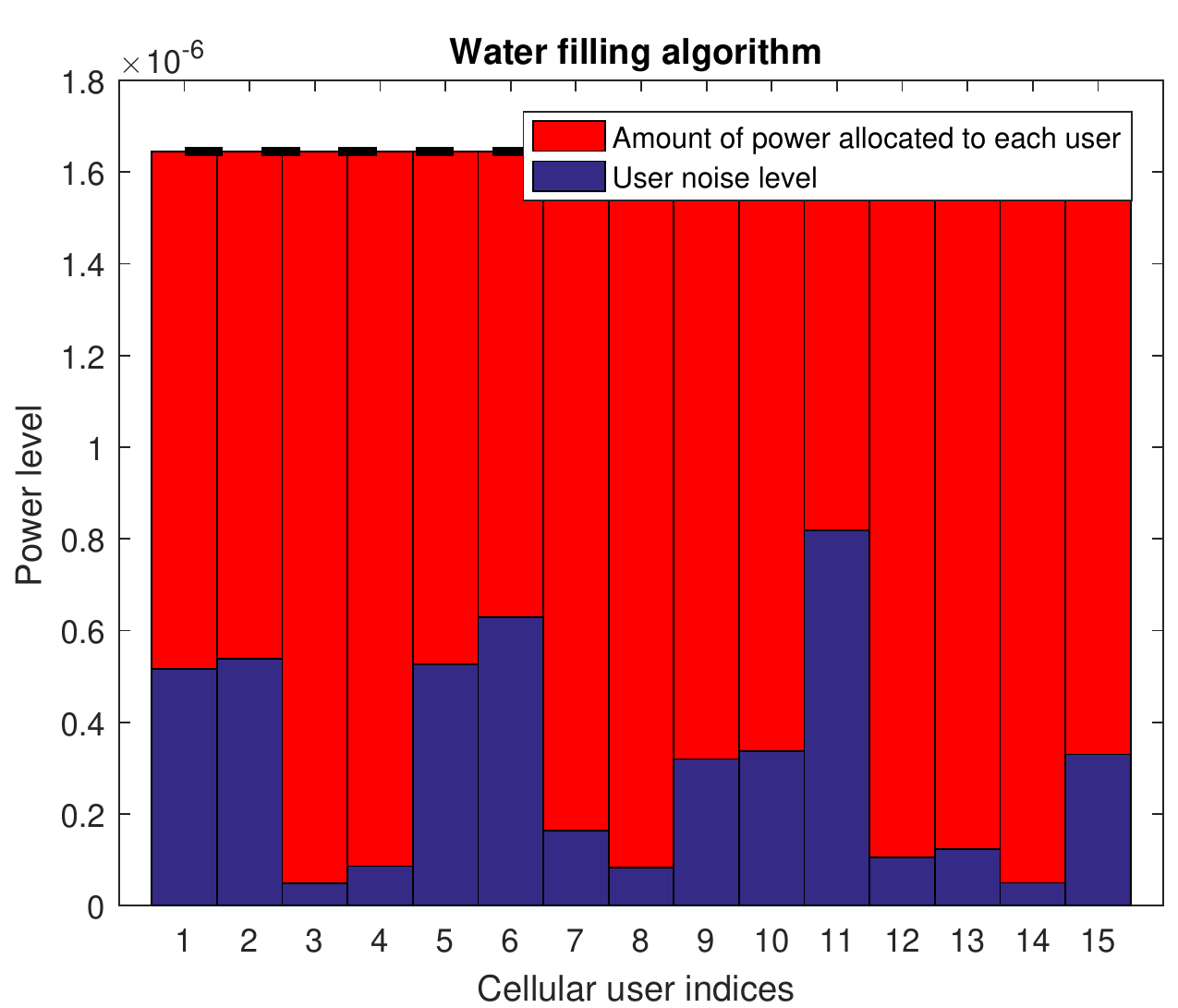}}\quad	
		\subfigure[\label{WF2}]{\includegraphics[width=0.3\linewidth]{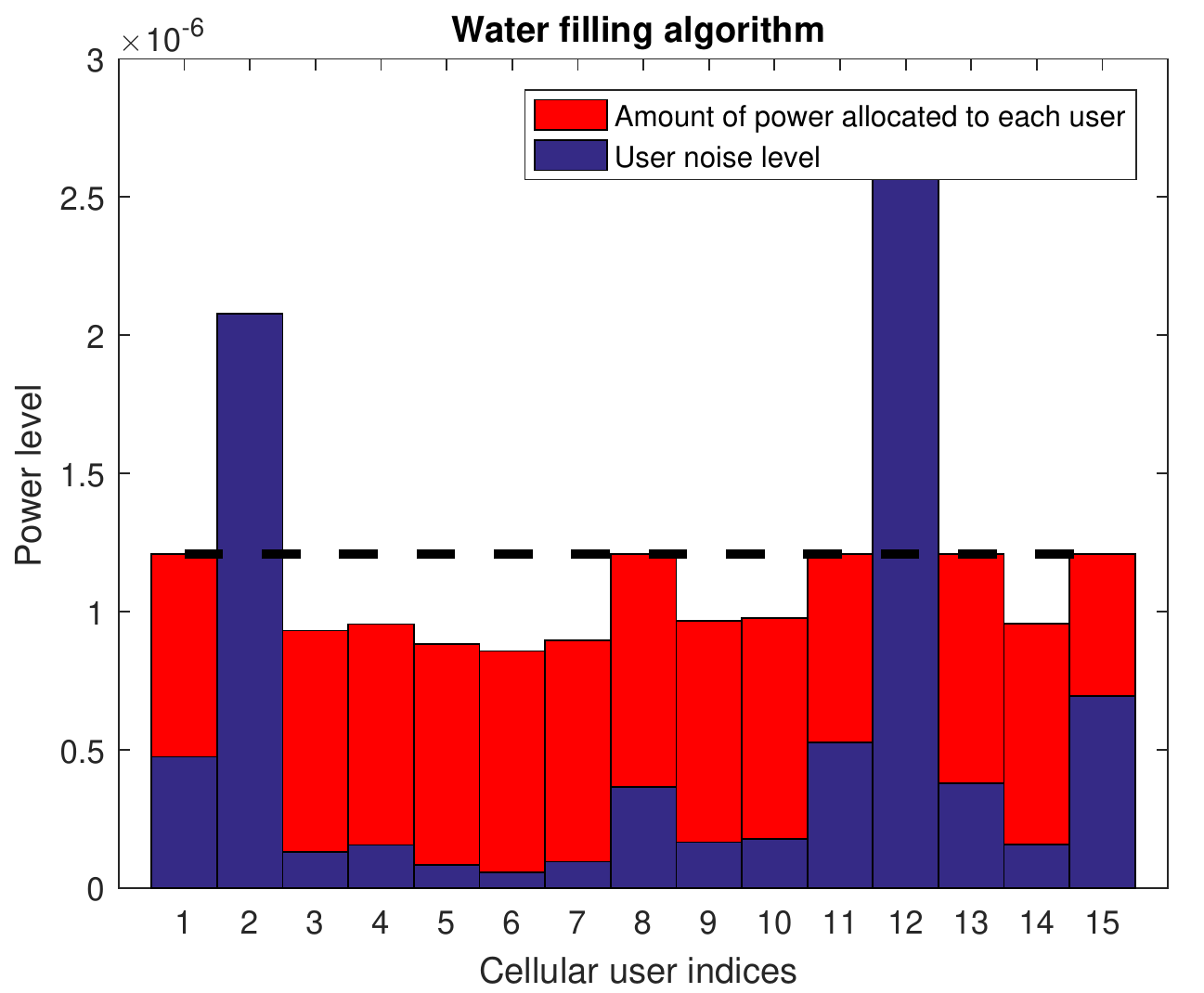}}\quad
		\subfigure[\label{WF3}]{\includegraphics[width=0.3\linewidth]{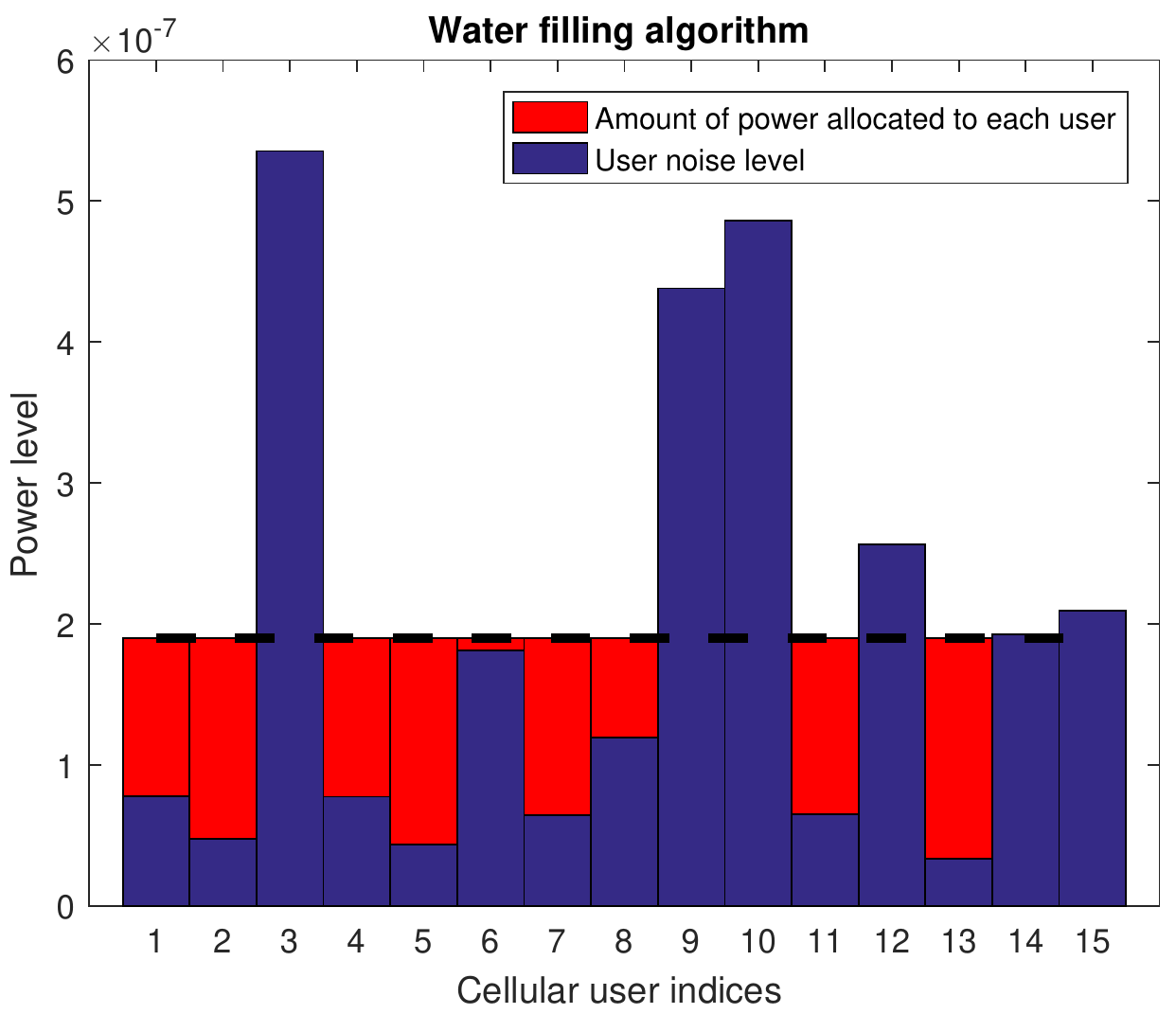}}	
		\caption{\ref{WF1} shows the sufficient power case where all users have good channel gain. \ref{WF2} shows the use case where the high channel gain users are allocated maximum power level $p_\textrm{max}$ and the low channel gain users are allocated power proportionally. \ref{WF3} shows the use case when all the users have very low channel gain as compared to the available power which is not sufficient for all the users. Therefore, only high channel gain users are allocated power. }	
		\label{results}
	\end{figure*}
	
	Fig. \ref{results} illustrates three possible cases of the water-filling algorithm for power allocation. We show the results for a different amount of powers to be allocated to the associated cellular users. In the first case Fig. \ref{WF1}, there is sufficient available power which is allocated to the associated users based on the proportional noise level. In the second case Fig. \ref{WF2}, there is enough power available that $p_\textrm{max}$ is allocated to the low-noise users and high-noise users are allocated power based on the water level. In the third case Fig. \ref{WF3}, the available power is insufficient; therefore, the power is allocated to each user based on the noise level while satisfying the water level feasibility.     
	
	\section{Conclusion}
	\label{secCon}
	
	In this paper, we have studied the UAV-assisted cellular networks to enhance the cellular network capacity. We have formulated a joint optimization problem for the user association and power allocation for the 5G NR traffic classifications. First, by utilizing the probability of ruin to estimate the possible number of cellular users to be associated with each UAV, we have solved the user association problem. Based on the probability of ruin of the UAV, the cellular traffic was offloaded by associating the cellular users with the other UAVs or the terrestrial network. Then, we have iteratively solved the power allocation problem. Simulation results have demonstrated the effectiveness of the proposed ruin-based energy-efficiency scheme. 

	\appendices
	\section{Proof of equivalence of (\ref{eq:obj1}) and (\ref{eq:obj2})}
	\label{appProofEqui}
	(\ref{eq:p1const1}) is separated for the UBSs $j \in \mathcal{U}$ as follows: 
	\begin{equation}
	\sum\limits_{k \in \mathcal{K}} P_{jk}\leq \rho_j , \quad \forall j \in \mathcal{U} ,
	\label{eq:proof1}
	\end{equation}
	where $\rho_j = \rho_0 + \rho t$ denoted the power level of UAV $j$. By rearranging (\ref{eq:proof1}), we get that
	\begin{equation}
	\rho_0 + \rho t - \sum\limits_{k \in \mathcal{K}} P_{jk} \geq 0 , \quad \forall j \in \mathcal{U}.
	\label{eq:proof2}
	\end{equation}
	Inequality (\ref{eq:proof2}) refers to the positive surplus power for every UAV which is the definition of the probability of ruin. Therefore, inequality (\ref{eq:proof2}) is equivalent to minimizing the probability of ruin. 
	
	\section{Proof of power allocation to URLLC users}
	\label{appProofChance}
	From \eqref{eq:p2const2}, $\textrm{Pr} \left( \gamma_{jk^{'}} \geq \zeta \right)$ can be expressed as CDF $F_{\gamma_{jk^{'}}} (\zeta )$. So we get:
	
	\begin{equation}
	F_{\gamma_{jk^{'}}} (\zeta ) \geq (1-\epsilon),
	\end{equation}
	From here, we can easily get the following expression for $\beta_k$
	
	\begin{equation}
	\zeta \geq \mathcal{F}^{-1}_{\gamma_{jk^{'}}} (1-\epsilon)
	\label{cdf}
	\end{equation}
	
	The maximization of the eMBB data rate under the constraint (\ref{eq:p2const2}) gives boundary solution $\gamma_{jk^{'}} = \zeta$ for the optimal power allocated to the URLLC users.	The optimal power is given as follows: 
	\begin{equation}
	P^{*}_{jk^{'}} = \frac{ \mathcal{F}^{-1}_{\gamma_{jk^{'}}} (1-\epsilon) (1 + I )}{h_{jk}},
	\label{ualloc1}
	\end{equation}
	where $I = \sum\limits_{j' \in \mathcal{J} \setminus \{0,j\} } P_{j'k}h_{j'k} + \omega_{jk} \sigma^2$.
	
	\section{Proof of equivalence of (\ref{eq:obj4}) and (\ref{eq:obj3})}
	\label{secproof2}
	
	The objective function of problem (\ref{eq:obj4}) is given as follows: 
	\begin{equation}
	\varsigma \sum\limits_{j \in \mathcal{J}} \sum\limits_{k \in \mathcal{K}}  D^{'}_{jk} - \xi \sum\limits_{u \in \mathcal{U}}\psi_u(\rho_0,t).
	\end{equation}
	It can be seen that the factor, $\xi \sum_{u \in \mathcal{U}}\psi_u(\rho_0,t)$ is constant when observed from the perspective of power allocation and therefore can be ignored in the optimization problem (\ref{eq:obj3}). Then after the URLLC association $x_{jk^{'}}$ according to Proposition 1 and the ruin-based eMBB association $x_{jk}$, the eMBB data rate is given as follows:
	
	\begin{equation}
	R^{'}_{jk} = \left(T-t \lambda_u \right) x^{*}_{jk}\omega_{jk} \log\left( 1 + \gamma_{jk}\right),
	\end{equation} 
	which is the objective function of the problem (\ref{eq:obj3}). 
	
	\section{Proof of Water Filling Algorithm using KKT}
	\label{secproof1}
	
	The optimization problem in (\ref{eq:obj3}) can be expressed in standard form as follows: 
	\begin{subequations}\label{eq:obj6}
		\begin{align}
		\min_{\boldsymbol{P}} \tag{\ref{eq:obj6}} \quad 
		& -\sum\limits_{j \in \mathcal{J}} \sum\limits_{k \in \mathcal{K}_e}  x^{*}_{jk}\omega_{jk} \log\left( 1 + \gamma_{jk}\right),  \\
		\text{s.t.}\quad 	
		&\label{eq:pconst1} \sum\limits_{k \in \mathcal{K}_e} P_{jk} = \rho_j - \sum\limits_{k^{'} \in \mathcal{K}_u} P^{*}_{jk^{'}} , \quad \forall j \in \mathcal{J} , \\	
		&\label{eq:pconst2} -P_{jk} \leq 0 , \quad \forall j \in \mathcal{J}, k \in \mathcal{K}_e \\
		&\label{eq:pconst3} P_{jk} \leq p_{\textrm{max}} , \quad \forall j \in \mathcal{J}, k \in \mathcal{K}_e.
		\end{align}
	\end{subequations}
	
	The Lagrangian function is given by 
	\begin{equation}
	\begin{aligned}
	\mathcal{L}( \boldsymbol{P},\boldsymbol{\lambda},\boldsymbol{\mu}, \boldsymbol{\nu}) = -\sum\limits_{j \in \mathcal{J}} \sum\limits_{k \in \mathcal{K}_e}  x^{*}_{jk}\omega_{jk} \log\left( 1 + \gamma_{jk}\right) \\ 
	+ \sum\limits_{j \in \mathcal{J}} \lambda_j \left(\sum\limits_{k \in \mathcal{K}_e} P_{jk} - \rho_j + \sum\limits_{k^{'} \in \mathcal{K}_u} P^{*}_{jk^{'}}\right) \\ 
	+ \sum\limits_{j \in \mathcal{J}} \sum\limits_{k \in \mathcal{K}_e}\mu_{jk} P_{jk} + \sum\limits_{j \in \mathcal{J}} \sum\limits_{k \in \mathcal{K}_e}\nu_{jk} (P_{jk}-p_{\textrm{max}}). 
	\end{aligned}
	\end{equation}
	
	In addition to the necessary conditions of problem (\ref{eq:obj6}), we obtain the following KKT conditions: 
	
	\begin{equation}
	\begin{aligned}
	\nabla \mathcal{L} (\boldsymbol{P})  =  - \frac{x^{*}_{jk}\omega_{jk}\theta_{jk}}{(1 + \theta_{jk} P_{jk}) } + \lambda_j - \mu_{jk} \\+ \nu_{jk} =0, \quad \forall j \in \mathcal{J}, k \in \mathcal{K}_e,
	\label{pkkt1}
	\end{aligned}
	\end{equation}
	
	\begin{equation}
	\mu_{jk} P_{jk} = 0,  \quad \forall j \in \mathcal{J}, k \in \mathcal{K}_e,
	\end{equation}
	
	\begin{equation}
	\nu_{jk} (P_{jk}-p_{\textrm{max}}),  \quad \forall j \in \mathcal{J}, k \in \mathcal{K}_e,
	\label{pkkt3}
	\end{equation}
	
	\begin{equation}
	\mu_{jk}, \nu_{jk} \geq 0 , \quad \forall j \in \mathcal{J}, k \in \mathcal{K}_e,
	\end{equation}
	
	$P_{jk} > 0 \implies \mu_{jk} = 0$. By concurrently solving (\ref{eq:pconst1}), (\ref{pkkt1}), and (\ref{pkkt3}), we have that 
	
	\begin{equation}
	P_{jk}^{*} = \min \left\{p_{\textrm{max}},
	\left[\frac{ x^{*}_{jk}\omega_{jk}}{\lambda_j}  - \frac{1}{\theta_{jk}} \right]^+ \right\} , \quad \forall j \in \mathcal{J}, k \in \mathcal{K}_e.
	\label{wtfill1}
	\end{equation}

	%
	%

	\bibliographystyle{IEEEtran}
	\bibliography{refUAV}

\begin{thebibliography}{10}
\providecommand{\url}[1]{#1}
\csname url@samestyle\endcsname
\providecommand{\newblock}{\relax}
\providecommand{\bibinfo}[2]{#2}
\providecommand{\BIBentrySTDinterwordspacing}{\spaceskip=0pt\relax}
\providecommand{\BIBentryALTinterwordstretchfactor}{4}
\providecommand{\BIBentryALTinterwordspacing}{\spaceskip=\fontdimen2\font plus
\BIBentryALTinterwordstretchfactor\fontdimen3\font minus
  \fontdimen4\font\relax}
\providecommand{\BIBforeignlanguage}[2]{{%
\expandafter\ifx\csname l@#1\endcsname\relax
\typeout{** WARNING: IEEEtran.bst: No hyphenation pattern has been}%
\typeout{** loaded for the language `#1'. Using the pattern for}%
\typeout{** the default language instead.}%
\else
\language=\csname l@#1\endcsname
\fi
#2}}
\providecommand{\BIBdecl}{\relax}
\BIBdecl

\bibitem{1677946}
R.~W. {Beard}, T.~W. {McLain}, D.~B. {Nelson}, D.~{Kingston}, and
  D.~{Johanson}, ``Decentralized cooperative aerial surveillance using
  fixed-wing miniature {UAVs},'' \emph{Proceedings of the IEEE}, vol.~94,
  no.~7, pp. 1306--1324, July 2006.

\bibitem{mozaffari2019beyond}
M.~{Mozaffari}, A.~{Taleb Zadeh Kasgari}, W.~{Saad}, M.~{Bennis}, and
  M.~{Debbah}, ``{Beyond 5G with UAVs: Foundations of a 3D wireless cellular
  network},'' \emph{IEEE Transactions on Wireless Communications}, vol.~18,
  no.~1, pp. 357--372, Jan 2019.

\bibitem{mozaffari2018tutorial}
M.~{Mozaffari}, W.~{Saad}, M.~{Bennis}, Y.~{Nam}, and M.~{Debbah}, ``{A
  tutorial on UAVs for wireless networks: Applications, challenges, and open
  problems},'' \emph{IEEE Communications Surveys Tutorials}, pp. 1--1, 2019.

\bibitem{saad2019vision}
W.~{Saad}, M.~{Bennis}, and M.~{Chen}, ``A vision of {6G} wireless systems:
  Applications, trends, technologies, and open research problems,'' \emph{IEEE
  Network}, to appear, 2020.

\bibitem{kalantari2017backhaul}
E.~Kalantari, M.~Z. Shakir, H.~Yanikomeroglu, and A.~Yongacoglu,
  ``Backhaul-aware robust {3D} drone placement in {5G+} wireless networks,'' in
  \emph{Proc. of the IEEE International Conference on Communications Workshops
  (ICC Workshops)}, Paris, France, May 2017.

\bibitem{challita2018cellular}
U.~{Challita}, W.~{Saad}, and C.~{Bettstetter}, ``Interference management for
  cellular-connected {UAVs}: A deep reinforcement learning approach,''
  \emph{IEEE Transactions on Wireless Communications}, vol.~18, no.~4, pp.
  2125--2140, April 2019.

\bibitem{saadmobsky}
R.~{Amer}, W.~{Saad}, and N.~{Marchettic}, ``Mobility in the sky: Performance
  and mobility analysis for cellular-connected {UAVs},'' \emph{IEEE
  Transactions on Communications}, to appear, 2020.

\bibitem{chen2017optimal}
J.~{Chen} and D.~{Gesbert}, ``Optimal positioning of flying relays for wireless
  networks: A {LOS} map approach,'' in \emph{Proc. of the IEEE International
  Conference on Communications (ICC)}, Chengdu, China, May. 2017.

\bibitem{sharma2016uav}
V.~{Sharma}, M.~{Bennis}, and R.~{Kumar}, ``{UAV}-assisted heterogeneous
  networks for capacity enhancement,'' \emph{IEEE Communications Letters},
  vol.~20, no.~6, pp. 1207--1210, June 2016.

\bibitem{uavhetnet}
A.~Merwaday and I.~Guvenc, ``{UAV} assisted heterogeneous networks for public
  safety communications,'' in \emph{Proc. of the IEEE Wireless Communications
  and Networking Conference Workshops (WCNCW)}, New Orleans, LA, USA, March
  2015.

\bibitem{saadtransport}
M.~Mozaffari, W.~Saad, M.~Bennis, and M.~Debbah, ``Wireless communication using
  unmanned aerial vehicles ({UAVs}): Optimal transport theory for hover time
  optimization,'' \emph{IEEE Transactions on Wireless Communications}, vol.~16,
  no.~12, pp. 8052--8066, Dec 2017.

\bibitem{khoshkholgh2019coverage}
M.~Khoshkholgh, K.~Navaie, H.~Yanikomerogluy, V.~Leung, K.~Shin \emph{et~al.},
  ``Coverage performance of aerial-terrestrial hetnets,'' \emph{arXiv preprint
  arXiv:1902.08547}, 2019.

\bibitem{uavhetnet1}
J.~Plachy, Z.~Becvar, P.~Mach, R.~Marik, and M.~Vondra, ``Joint positioning of
  flying base stations and association of users: Evolutionary-based approach,''
  \emph{IEEE Access}, vol.~7, pp. 11\,454--11\,463, Jan 2019.

\bibitem{bennis2018ultrareliable}
M.~Bennis, M.~Debbah, and H.~V. Poor, ``Ultrareliable and low-latency wireless
  communication: Tail, risk, and scale,'' \emph{Proceedings of the IEEE}, vol.
  106, no.~10, pp. 1834--1853, Oct 2018.

\bibitem{saadiot}
M.~Mozaffari, W.~Saad, M.~Bennis, and M.~Debbah, ``Mobile unmanned aerial
  vehicles ({UAVs}) for energy-efficient {Internet of Things} communications,''
  \emph{IEEE Transactions on Wireless Communications}, vol.~16, no.~11, pp.
  7574--7589, Nov 2017.

\bibitem{fadlullah2016dynamic}
Z.~M. Fadlullah, D.~Takaishi, H.~Nishiyama, N.~Kato, and R.~Miura, ``A dynamic
  trajectory control algorithm for improving the communication throughput and
  delay in {UAV}-aided networks,'' \emph{IEEE Network}, vol.~30, no.~1, pp.
  100--105, Jan 2016.

\bibitem{ren2019achievable}
H.~Ren, C.~Pan, K.~Wang, Y.~Deng, M.~Elkashlan, and A.~Nallanathan,
  ``Achievable data rate for {URLLC}-enabled {UAV} systems with {3-D} channel
  model,'' \emph{arXiv preprint arXiv:1907.06557}, 2019.

\bibitem{pan2019joint}
C.~Pan, H.~Ren, Y.~Deng, M.~Elkashlan, and A.~Nallanathan, ``Joint blocklength
  and location optimization for {URLLC}-enabled {UAV} relay systems,''
  \emph{IEEE Communications Letters}, vol.~23, no.~3, pp. 498--501, March 2019.

\bibitem{maggreen}
S.~Koulali, E.~Sabir, T.~Taleb, and M.~Azizi, ``A green strategic activity
  scheduling for {UAV} networks: A sub-modular game perspective,'' \emph{IEEE
  Communications Magazine}, vol.~54, no.~5, pp. 58--64, May 2016.

\bibitem{ruieerelay}
J.~Zhang, Y.~Zeng, and R.~Zhang, ``Spectrum and energy efficiency maximization
  in {UAV}-enabled mobile relaying,'' in \emph{Proc. of the IEEE International
  Conference on Communications (ICC)}, Chengdu, China, May 2017.

\bibitem{ee}
L.~Zhang, Z.~Zhao, Q.~Wu, H.~Zhao, H.~Xu, and X.~Wu, ``Energy-aware dynamic
  resource allocation in {UAV} assisted mobile edge computing over social
  internet of vehicles,'' \emph{IEEE Access}, vol.~6, pp. 56\,700--56\,715, Oct
  2018.

\bibitem{alzenad20173}
M.~Alzenad, A.~El-Keyi, F.~Lagum, and H.~Yanikomeroglu, ``{3-D} placement of an
  unmanned aerial vehicle base station ({UAV-BS}) for energy-efficient maximal
  coverage,'' \emph{IEEE Wireless Communications Letters}, vol.~6, no.~4, pp.
  434--437, Aug 2017.

\bibitem{zeng2017energy}
Y.~Zeng and R.~Zhang, ``Energy-efficient {UAV} communication with trajectory
  optimization,'' \emph{IEEE Transactions on Wireless Communications}, vol.~16,
  no.~6, pp. 3747--3760, June 2017.

\bibitem{yang2019energy}
Z.~Yang, C.~Pan, K.~Wang, and M.~Shikh-Bahaei, ``Energy efficient resource
  allocation in {UAV}-enabled mobile edge computing networks,'' \emph{arXiv
  preprint arXiv:1902.03158}, 2019.

\bibitem{harvestalsharoaUAV}
A.~{Alsharoa}, H.~{Ghazzai}, A.~{Kadri}, and A.~E. {Kamal}, ``Spatial and
  temporal management of cellular hetnets with multiple solar powered drones,''
  \emph{IEEE Transactions on Mobile Computing}, vol.~19, no.~4, pp. 954--968,
  April 2020.

\bibitem{long2018energy}
T.~Long, M.~Ozger, O.~Cetinkaya, and O.~B. Akan, ``Energy neutral internet of
  drones,'' \emph{IEEE Communications Magazine}, vol.~56, no.~1, pp. 22--28,
  Jan 2018.

\bibitem{sekander2019performance}
S.~Sekander, H.~Tabassum, and E.~Hossain, ``On the performance of renewable
  energy-powered {UAV}-assisted wireless communications,'' \emph{arXiv preprint
  arXiv:1907.07158}, 2019.

\bibitem{yang2018outage}
L.~Yang, J.~Chen, M.~O. Hasna, and H.-C. Yang, ``Outage performance of
  {UAV}-assisted relaying systems with {RF} energy harvesting,'' \emph{IEEE
  Communications Letters}, vol.~22, no.~12, pp. 2471--2474, Dec 2018.

\bibitem{vincent2018prospects}
D.~Vincent, P.~S. Huynh, L.~Patnaik, and S.~S. Williamson, ``Prospects of
  capacitive wireless power transfer ({C-WPT}) for unmanned aerial vehicles,''
  in \emph{Proc. of the IEEE PELS Workshop on Emerging Technologies: Wireless
  Power Transfer (Wow)}.\hskip 1em plus 0.5em minus 0.4em\relax Montréal, QC,
  Canada: IEEE, June 2018.

\bibitem{finiteruin}
W.-S. Chan and L.~Zhang, ``Direct derivation of finite-time ruin probabilities
  in the discrete risk model with exponential or geometric claims,''
  \emph{North American Actuarial Journal}, vol.~10, no.~4, pp. 269--279, Oct
  2006.

\bibitem{davis2005insurance}
M.~Davis, J.~Hylands, S.~Life, J.~McCutcheon, R.~Norberg, H.~Panjer, A.~Wilson,
  and W.~Wyatt, ``Insurance risk and ruin,'' Oct 2005.

\bibitem{mozaffari2016efficient}
M.~{Mozaffari}, W.~{Saad}, M.~{Bennis}, and M.~{Debbah}, ``Efficient deployment
  of multiple unmanned aerial vehicles for optimal wireless coverage,''
  \emph{IEEE Communications Letters}, vol.~20, no.~8, pp. 1647--1650, Aug 2016.

\bibitem{3gpp.36.331}
3GPP, ``{Evolved Universal Terrestrial Radio Access (E-UTRA); Radio Resource
  Control (RRC); Protocol specification},''
  \url{{https://portal.3gpp.org/desktopmodules/Specifications/SpecificationDetails.aspx?specificationId=2440}},
  {3rd Generation Partnership Project (3GPP)}, Technical Specification (TS)
  36.331, 04 2017, version 14.2.2.

\bibitem{saadicc}
U.~Challita, W.~Saad, and C.~Bettstetter, ``Deep reinforcement learning for
  interference-aware path planning of cellular-connected {UAVs},'' in
  \emph{Proc. of the IEEE International Conference on Communications (ICC)},
  Kansas City, MO, USA, May 2018.

\bibitem{aunasruin}
A.~Manzoor, N.~H. Tran, W.~Saad, S.~A. Kazmi, S.~R. Pandey, and C.~S. Hong,
  ``Ruin theory for dynamic spectrum allocation in {LTE-U} networks,''
  \emph{IEEE Communications Letters}, vol.~23, no.~2, pp. 366--369, Dec 2018.

\bibitem{poisson}
F.~{Qamar}, K.~{Dimyati}, M.~N. {Hindia}, K.~A. {Noordin}, and I.~S. {Amiri},
  ``A stochastically geometrical {Poisson} point process approach for the
  future {5G D2D} enabled cooperative cellular network,'' \emph{IEEE Access},
  vol.~7, pp. 60\,465--60\,485, May 2019.

\end{thebibliography}

\end{document}